\documentclass[12pt, draftclsnofoot, onecolumn]{IEEEtran}
\usepackage{graphicx} % Allows including images
\usepackage{booktabs} % Allows the use of \toprule, \midrule and \bottomrule in tables
\usepackage{amsmath}
\usepackage{mathrsfs}
\usepackage{amsfonts}
\usepackage{amssymb}
\usepackage{epstopdf}
\usepackage{dblfloatfix}
\usepackage[caption=false,font=footnotesize]{subfig}
\usepackage{color}
\usepackage{enumitem}
\usepackage{amsthm}
\usepackage{algorithm,algorithmic}
\usepackage{multirow}
\usepackage{array}
\usepackage{cite}
\usepackage{color}
\usepackage{hyperref}

\theoremstyle{remark}

\theoremstyle{definition}

\newcommand{\R}{\mathbb{R}}

\newcommand{\DIF}{\mathrm{dif}}

\title{Wireless Channel Modeling Perspectives for Ultra-Reliable Low Latency Communications}
\author{\IEEEauthorblockN{Patrick C. F. Eggers, Marko Angjelichinoski, and Petar Popovski} \\
Department of Electronic Systems, Aalborg University, Denmark\\
}
\begin{document}
\maketitle
\begin{abstract}
Ultra-Reliable Low Latency Communication (URLLC) is one of the distinctive features of the upcoming 5G wireless communication, going down to packet error rates (PER) of $10^{-9}$. In this paper we discuss the statistical properties of the wireless channel models that are relevant for characterization of the lower tail of the  Cumulative Distribution Function (CDF). 
We show that, for a wide range of channel models, the outage probability at URLLC levels can be approximated by a simple power law expression, whose exponent and offset depend on the actual channel model.
The main insights from the analysis can be summarized as follows: (1) the two-wave model and the impact of shadowing in combined models lead to pessimistic predictions of the fading in the URLLC region; (2) the CDFs of models that contain single cluster diffuse components have slopes that correspond to the slope of a Rayleigh fading, and (3) multi-cluster diffuse components can result in different slopes.
We apply our power law approximation results to analyze the performance of receiver diversity schemes for URLLC-relevant statistics and obtain a new simplified expression for Maximum Ratio Combining (MRC) in channels with power law tail statistics.
% * <pe@es.aau.dk> 2017-11-28T16:30:31.670Z:
%
% ^.
\end{abstract}
\begin{IEEEkeywords} 
Ultra-reliable communications, URLLC, 5G, wireless channel models, fading, diversity, probability tail approximations, rare event statistics. 
\end{IEEEkeywords}

\IEEEpeerreviewmaketitle

\section{Introduction} % (fold)
\label{sec:introduction}

One of the features of 5G wireless communication systems is to offer
wireless communication with extremely high reliability guarantees,
also known as \emph{Ultra-Reliable Low Latency Communication (URLLC)} \cite{PetarURC,PetarURC2}.
The level of reliability, going sometimes down to packet error rates (PER)
of $10^{-9}$, should be sufficiently convincing in order to remove
cables in an industrial setting, remote control of robots and drones
that need to perform a critical function, remote surgery or self-driving
cars \cite{schulz2017latency}. A significant number of the URLLC use cases take
place indoors, which resonates with the expectation for a wireless environment that is, to some extent, controlled and offers
predictable signal quality. Wireless communication performance
is inherently a stochastic variable; however, URLLC refocuses our attention
to the statistical characterization of the rare events and the lower
tail of the channel Cumulative Distribution Function (CDF). This requires
a major effort in terms of measurement campaigns, purposefully designed
to capture the lower tail statistics, as well as in terms of analytical
channel modeling.

Nevertheless, a good starting point for characterizing wireless channels
is to analyze the current channel models  in regimes that are relevant
for URLLC, which is the topic of this article. The related works  \cite{Durgin2002,rao2015mgf,Yoo2016,Yoo2015,Paris2014} have characterized the probability density functions of wireless channel models, often leading to very complex expression and analysis dependent on multiple parameters. 
Our objective is to get
insight into the behavior of different channel models at the region of \emph{ultra-low error rates}, which we define as a PER that is $\epsilon\leq10^{-5}$ or lower, corresponding to the reliability of ``five nines''. We have selected $10^{-5}$ as this is the target error rate for URLLC selected in 3GPP for a packet of $32$ bytes to be delivered within 1 ms \cite{3gpp38.913}. Different mission-critical services will use different levels of ultra-reliability, such as PERs of $10^{-6}$ in smart grids and $10^{-9}$ for factory automation~\cite{schulz2017latency}. We focus on the packet errors that occur due to outages, induced by block fading, rather than errors caused by noise. Recent studies \cite{durisi2016toward} have shown that this is a very suitable model for transmission of short packets, which are in turn expected to be prevalent in the URLLC scenarios. Throughout the paper we will use the term \emph{URLLC-relevant statistics} to denote the events in signal reception that occur with probabilities on the order of $10^{-5}$ or less. Our analysis shows that, despite the complex probability density functions, the behavior of the lower tail can be significantly simplified, leading to important insights on the behaviors that occur when the outage probabilities are very low. While the tuning of the models can only be done with appropriate experimental evidence, this work provides a valuable contribution to the modeling efforts that are essential for designing ultra-reliable wireless systems.  
% * <pe@es.aau.dk> 2017-11-29T16:20:39.852Z:
%
% ^.
%Furthermore, it is confirmed that the two-wave model with equal amplitudes of the two waves represents one of the most pessimistic cases in terms of URC-relevant statistics. 
% * <pe@es.aau.dk> 2017-11-28T16:36:26.144Z:
%
% ^.

Besides the analysis of various statistical channel models, in this paper we also revisit the basic assumptions
adopted in channel modeling in the light of the statistically rare events that needs to be captured in an URLLC regime. The existing channel models have been developed for wireless communication systems\footnote{For example, the first generation digital systems, such as GSM, that had an emphasis on voice communication.} that deal with bit error rates (BER) of $10^{-3}$ to $10^{-4}$ {\cite{etsi_gsm_standard2014} and there is a lack of experimental evidence to support channel models that deal with PER at the URLLC levels. This paper extrapolates the standard channel models towards URLLC-relevant statistics and identifies their limitations when modeling rare events. One of our main result is that, for a wide range of channel models (but not all, as it will be seen in the paper),  
% * <pe@es.aau.dk> 2017-11-29T16:22:15.395Z:
%
% ^.
the tail approximation of the cumulative distribution function at the URLLC levels has the form:
\begin{equation}\label{eq:MainResult}
F\left(\frac{P_R}{A}\right)\approx \alpha \left(\frac{P_R}{A}\right)^{\beta}
\end{equation}
where $A$ is the average power, $P_R$ is the minimal required power to decode the packet correctly and $\alpha, \beta$ are parameters that depend on the actual channel model.
%Specifically, the exponent $\beta$ determines how the error probability depends on the minimal required power.
% * <pe@es.aau.dk> 2017-12-01T16:32:20.153Z:
%
% ^.

Since the support of URLLC will be significantly dependent on high levels of diversity, we have also characterised the URLLC-relevant statistics when multiple antennas are considered. Specifically, we provide a simplified analysis of $M$-branch receive diversity for uncorrelated branch signals, that makes use of (\ref{eq:MainResult}), as well as approximations for some special channels. The result provides a compact Maximum Ratio Combing (MRC) solution of form
\begin{equation}
F_{\mathrm{MRC}} \approx \alpha_{\mathrm{MRC}}(\beta_1..\beta_M)F_{\mathrm{SC}}
\end{equation}
that is, a scaled version of a Selection Combining (SC) solution, in which the scale parameter $ \alpha_{\mathrm{MRC}}$ depends only on the branch exponents $\beta$.
% * <m.angjelichinoski@gmail.com> 2018-01-02T22:32:56.836Z:
%
% ^.

The paper is organized as follows.
% * <pe@es.aau.dk> 2017-11-28T16:55:52.182Z:
%
% ^.
After the introduction, in Section II we provide the system model. Section III contains analysis of a wide range of channel models which exhibit power law tails at URLLC-relevant probabilities. % Also approximation error analysis is given for the more fundamental models.
Section IV contains analysis of two models that do not not result in power law tails. Section V contains the analysis of the receive diversity schemes in URLLC-relevant regime. Section VI concludes the paper.

\section{Wireless Channel Modeling for URLLC-Relevant Statistics}
\label{sec:sysmodel}

\subsubsection{Preliminaries}
The common approach in wireless channel modeling is to assume \emph{separability} of the following effects~\cite{RVJBA}:
\begin{itemize}
\item {Path loss}, dependent on the actual geometric setting and operating frequency. 
\item Long-term fading (i.e., {shadowing}) that captures slowly-varying macroscopic effects. 
\item Short-term fading processes, relevant on a time scale of a packet (i.e., quasi-static fading) or even a symbol (fast fading), assuming stationary scattering conditions.
\end{itemize}
The URLLC performance is determined by the short-term process and its (un)predictability, which ultimately determines the fate of the packet at the destination.
Assuming separability, the statistics of short-term fading is described via parameters that are derived from the long-term fading and path loss effects; these parameters are assumed to be constant over a period of time.
However, separability becomes problematic when URLLC-relevant statistics is considered, since the estimated long-term parameters require certain level of accuracy in order to have a valid short-term statistics of rare events. Motivated by this, we also consider \emph{combined} long and short term fading models. Furthermore, in absence of dedicated URLLC channel models, we investigate the behavior of a wide palette of existing wireless channel models in URLLC regime. 

\subsubsection{General Model}
We use combination of (a) the complex baseband model of a narrowband channel with reduced wave grouping from \cite{Durgin2002}, and (b) the incoherent multi-cluster channel of \cite{Nakagami1958,Yacoub2007}.
Let $P$ denote the total received power; we have:
\begin{equation} \label{eq:RWGmodel}
 P=\omega\sum_{m=1}^\mu |V_m|^{2/\beta},\; V_m=  \xi \left( \sum_{i=1}^N \rho_{i,m} e^{j\phi_{i,m}} \right)+ \prod_{l=1}^L V_{\mathrm{dif},m,l}. %^{1/\beta}%^{\tfrac{1}{2}}
\end{equation}
$V_m$ denotes the complex received voltage from the $m-$th cluster $m=1,\hdots,\mu$, in which $\rho_{i,m}/\phi_{i,m}$ is the amplitude/phase of the $i-$th specular component, $i=1,\hdots,N$ and $V_{\mathrm{dif},m,l}$ is the $l-$th diffuse component for the $m-$th cluster with $L$ denoting the number of diffuse components per cluster \cite{JBAIZK}.
%{The diffuse components are modeled using the multi-link feature \cite{JBAIZK}, such that $V_{\mathrm{dif},m,l}$ is the $l-$th diffuse component for the $m-$th cluster.} % In Cascaded Rayleigh channel $L=2$ \cite{chau2012second}, while $L=1$ for all other channels.
$\beta$ caters for the modeling of a Weibull channel \cite{bessate2016}, and for all other models  it is set to $\beta=1$.
%(each with same mean power $2\sigma^2$, see section YY). 
The shadowing effects are represented by the random variables (RVs) $\xi$ and $\omega$.
Here $\xi$ is a common shadowing amplitude that affects only the specular components \cite{Paris2014}, while $\omega$ induces a shadowing effect on the total power \cite{Yoo2015,RVJBA}, see section \ref{sec:generalizations}.
We assume that each $\rho_i$ of a specular component is constant and that $\phi_i$ is a uniform random variable \cite{Durgin2002}.
The elementary diffuse components $V_{\mathrm{dif},m,l}$ are treated in their simplest form, as a  contribution from a large number of waves and application of the central limit theorem ~\cite{Durgin2002,Nakagami1958,Yacoub2007}, which leads to $V_{\mathrm{dif}}=X_R+jX_I$, 
%\begin{equation}
%V_{\mathrm{dif}}=X_R+jX_I
%\end{equation}
where $X_R$ and $X_I$ are independent Gaussian variables, each with zero mean and variance $\sigma^2$. 
A more general variant of the diffuse component follows from a multi-scatter physical setup \cite{JBAIZK,Nakagami1958,Yacoub2007}. This leads to the cases of Nakagami, Weibull and Cascaded Rayleigh channel, as well as compound channels, such as Suzuki and shadowed $\kappa-\mu$ \cite{RVJBA,Yoo2016,Yoo2015,Paris2014}.

We treat narrowband channel models with block fading, such that the power at which the packet is received remains constant and equal to $P$ given with (\ref{eq:RWGmodel}). % and \MA{$r=\sqrt{P}$} is the envelope.
The noise power is normalized to $1$, such that $P$ also denotes the Signal-to-Noise Ratio (SNR) at which a given packet is received. For each new packet, all RVs from (\ref{eq:RWGmodel}) are independently sampled from their probability distributions\footnote{The reader may object that this assumption is not valid when long-term shadowing is treated, i.e. a sample for a given $\rho_i$ is applicable to several packet transmissions. See Section~\ref{sec:Lognormal} for discussion about this assumption.}. The average received power for the channel model (\ref{eq:RWGmodel}) across many realizations is denoted by $A$ and can be computed as:
\begin{equation}
A=E[P] \stackrel{(a)}{=} \overline{\omega}\sum_{m=1}^\mu \left[ \overline{\xi^2}\left(\sum_{i=1}^N \rho_{i,m}^2\right)+E \left[\prod_{l=1}^L |V_{\DIF, m, l}|^2 \right] \right],
\end{equation}
with $E[\cdot]$ denoting the expectation operator.
Note that (a) is valid when we treat the reduced wave grouping model from \cite{Durgin2002}.
In the subsequent analysis we assume normalized shadowing power, i.e. $\overline{\omega}=1$ and $\overline{\xi^2}=1$.
%where  
%$E[\cdot]$ denotes the expectation operator and (a) is valid when we treat the reduced wave grouping model from \cite{Durgin2002}. 
The diffuse term depends on link signal correlation, while for a single link ($L=1$) the average power of the elementary terms is $E[|V_{\DIF, m}|^2]=2\sigma^2$.

\subsubsection{Descriptive Metrics}
The specular component vector balancing in the reduced wave group model of \cite{Durgin2002} is given via the peak to average ratio of the two dominant specular powers: 
\begin{equation} \label{DeltaDefinition}
\Delta=\frac{2\rho_1\rho_2}{\rho_1^2+\rho_2^2}.
\end{equation}
Furthermore, the power ratio of the specular components and the diffuse component per cluster, called \emph{k-factor} is defined as $k_N=\frac{\sum_{i=1}^N \rho_i^2}{2\sigma^2}$, which in case of the multiple clusters gives \cite{Yacoub2007}: % gives:  
\begin{equation} \label{eq:k-factor}
\kappa=\frac{1}{\mu} \sum_{m=1}^{\mu} k_{N,m}= \frac{\sum_{m=1}^{\mu} \sum_{i=1}^N\rho_{i,m}^2}{\mu \cdot 2\sigma^2}.
 \end{equation}

\subsubsection{URLLC-Relevant Statistics}
Let $R$ denote the transmission rate of the packet. We assume that packet errors occur due to outage only, such that the PER $\epsilon $ is given by:
\begin{equation} \label{eq:DefPR}
\epsilon=\Pr(R<\log_2(1+P))=\Pr(P<P_R),
\end{equation}
where  $P_R=2^R-1$ is the minimal required power to receive the packet sent at rate $R$.
Denote by $\epsilon$ the target packet error probability (PER), also referred to as \emph{outage probability}. Then, for each model the objective is to find $P_R$, defined in (\ref{eq:DefPR}) through the CDF $F(P_R)$, obtained as:
\begin{equation}\label{eq:EpsilonIntegral}
\epsilon=F(P_R)=\int_{r_{\min}}^{\sqrt{P_R}} f(r) \mathrm{d}r,
\end{equation}
where $r=\sqrt{P}$ is the received envelope and $r_{\min}$ is the minimal value of the envelope in the support set of $f(r)$, which is the Probability Density Function (PDF) of the specific channel model. The key to the approximations presented in this paper is the fact that, for URLLC scenarios, $\epsilon$ is \emph{very small}.

\section{Channels with Power Law Tail Statistics}
We analyze the behavior of common wireless channel models in URLLC-relevant regime and derive asymptotically tight approximations $\tilde{\epsilon}$ of their tail probabilities $\epsilon$, that satisfy $ \lim_{P_R\rightarrow 0}\tilde{\epsilon}=\epsilon$
The common trait of all models considered in this section is that $\tilde{\epsilon}$ takes the form of a simple power law \eqref{eq:MainResult}. % or in log-domain:
%\begin{equation}\label{eq:log-form}
%\log \tilde{\epsilon} = \beta \log \left( \frac{P_R}{A} \right) + \log \alpha.
%\end{equation}
%and inversions as
%\begin{equation} \label{eq:inversion-form}
%P_R=A \left(\frac{\tilde{\epsilon}}{\alpha}\right)^{\frac{1}{\beta}},
%\end{equation}
with distribution-specific values of the parameters $\alpha, \beta$. % given in Table \ref{tab:tail_approx}.
\begin{table*}[t]
 \centering
  \caption{CDF tail approximations $\tilde{\epsilon}$, for distributions in the following subsections. %Validity power limits 
Relative power $p=P_R/A$, gain offset (scale) $\alpha$ and log-log slope (shape) $\beta=\tfrac{d\log\left(F\right)}{d\log\left(p\right)}$ following (\ref{eq:MainResult}).%  Bracketed entries $()^*$, indicate special solutions. 
  }
 \label{tab:tail_approx}
 \begin{tabular}{|l|c|c|c|c|c|}
 \hline
Channel Model & Tail $\tilde{\epsilon}=\lim_{p\rightarrow 0}F(p)$ &  %Limits & 
 Offset $\alpha$ & Slope $\beta$ \\
  \hline
 \hline
TW & $\frac{1}{2}-\frac{1}{\pi} \mathrm{asin} \left( \tfrac{1-p}{\Delta}\right)$ & %$1-\Delta\leq p \leq 1+\Delta$ & 
$ \frac{\sqrt{2}}{\pi}$ when $ \Delta \rightarrow 1$ & $\tfrac{1}{2}$ \\ 
 \hline
3W & $\frac{P_R}{4\pi\Delta_r}$ & %$\frac{r_{\min}^2}{A} \leq p \ll (\rho_1-\rho_2-\rho_3)^2/A$ &
$\frac{1}{4\pi\Delta_r}$ when $\rho_1<\rho_2+\rho_3 $ & 1  \\
 \hline

Rayl  & $p$ &  %$0\leq p \ll 1$ & 
1 & 1 \\
 \hline

Rice & $F_{\mathrm{Rayl}}\left(p(k_1+1)\right)e^{-k_1} $ &  %$0 \leq p \ll \tfrac{1}{4k_1(k_1+1)}$ & 
$(k_1+1)e^{-k_1}$ & 1 \\
 \hline

TWDP & $F_{\mathrm{Rice}}\left(p;k_2\right)I_{0}\left(k_{2}\Delta\right)$ & %$0 \leq p \ll \tfrac{1}{4k_2(k_2+1)}$ & 
$(k_2+1)e^{-k_2}I_{0}\left(k_{2}\Delta\right)$ & 1 \\
 \hline

Wei & ${\left(\Gamma(1+1/\beta)p\right)^{\beta}}$ & %$0 \leq p \ll 1/\Gamma(1+1/\beta)$ & 
$\Gamma(1+1/\beta)^\beta$ & $\beta$ \\
 \hline

Nak & $\left(m^{m}/\Gamma(m+1)\right)p^m$ & %$0 \leq p \ll \frac{1+1/m}{m}$  & 
$m^{m}/\Gamma(m+1)$ & $m$
 \\
 \hline

$\kappa\mu$ & $ F_{\mathrm{Nak}}(p;\mu) F_{\mathrm{Rice}}(1;\kappa)^\mu $ & %$0 \leq p \ll  \frac{\mu+1}{\mu(1+\kappa)}\frac{1}{\mu|1-\kappa|}$ & 
$\tfrac{\left( e^{-\kappa}(\kappa+1)\mu\right)^\mu}{\Gamma(\mu+1)}$ & $\mu$
 \\
 \hline

$\kappa\mu/m$ & $F_{\mathrm{Nak}}(p;\mu)(1+\kappa)^{\mu}\left(\frac{m}{\kappa\mu+m}\right)^m$& %$0 \leq p \ll \frac{\mu+1}{\mu(1+\kappa)}\frac{\kappa\mu+m}{|m\mu(1-\kappa)|+\kappa\mu^2}$ & 
$\frac{\mu^{\mu}(1+\kappa)^{\mu}}{\Gamma(\mu+1)}\left(\frac{m}{\kappa\mu+m}\right)^m$ & $\mu$
 \\
 \hline

$\kappa\mu/\alpha$ & $F_{\mathrm{\kappa\mu}}(p;\kappa,\mu)\cdot \frac{\Gamma(\alpha+\mu)}{(\alpha-1)^{\mu}\Gamma(\alpha)}$& %$0 \leq p \ll \frac{\mu+1}{\mu(1+\kappa)}| \frac{\alpha-1}{\kappa\mu(\alpha+\mu)-(\mu+1)}|$ & 
$\alpha_{\kappa\mu} \cdot \frac{\Gamma(\alpha+\mu)}{(\alpha-1)^{\mu}\Gamma(\alpha)}$ & $\mu$
 \\
 \hline

Suz & $P_R10^{\tfrac{1}{10} \left(\sigma_{\mathrm{dB}}^{2} \left( \frac{\ln10}{20} \right) -\mu_{\mathrm{dB}}\right)}$ &  %$0 \leq p \ll e^{-8\sigma_l^2}$& 
$10^{\tfrac{1}{10} \left(\sigma_{\mathrm{dB}}^{2} \left( \frac{\ln10}{20} \right) -\mu_{\mathrm{dB}}\right)}$ & 1 \\
 \hline

Cas & $- p \tfrac{1+\Gamma}{1-\Gamma}{\ln\left(p \tfrac{1+\Gamma}{(1-\Gamma)^2}\right)}$ & %$0 \leq p \ll \tfrac{1}{4}\tfrac{\left(1-\Gamma\right)^2}{1+\Gamma}$ & 
- & $1+\tfrac{1}{\ln \left( p\right)+\ln\tfrac{1+\Gamma}{(1-\Gamma)^2}}$ \\
 \hline

LN & $\tfrac{1}{4}e^{-\frac{\left(\frac{1}{2}\ln (P_R)-a\sigma_{l}-\mu_{l}\right)^{2}}{2\sigma_{l}^{2}}}$ & %$0 \leq p \ll e^{-2\sigma_l^2}$&
- & $\frac{10}{\ln10} \left[\frac{a}{\sigma_{\mathrm{dB}}}-2\frac{P_{R,\mathrm{dB}}-\mu_{\mathrm{dB}}}{2\sigma_{\mathrm{dB}}^{2}} \right]$ \\
\hline

  \end{tabular}
 \end{table*}

For the channel models that are important in practice (see subsections \ref{sec:TW}-\ref{sec:km}), we also provide a simple tool to analyze the convergence of $\tilde{\epsilon}$ to $\epsilon$. Specifically, we introduce the non-negative \emph{approximation error function} $\phi(P_R)$ that satisfies the inequality (see Appendix A):% accuracy of the tail approximation via the following inequality: % \PP{I am not sure how operational/easy to use is this inequality in the sequel of the text. For example, in the Rayleigh channel $\phi(P_R)$ is half of the value of the approximation. What does that mean for the error? Is it low, high? we need to defined a function that is more readily usable. In fact, characterizing $\frac{1}{1-\phi(P_R)}$ is better than only characterizing $\phi(P_R)$}:
\begin{equation} \label{eq:Def bounding error}
\tilde{\epsilon}(1-\phi(P_R))\leq\epsilon \leq\tilde{\epsilon}(1+\phi(P_R)).
\end{equation}
$\phi(P_R)$ increases monotonically with $P_R$ and satisfies $\lim_{P_R\rightarrow 0}\phi(P_R) = 0$.
We say that $\tilde{\epsilon}$ converges asymptotically to ${\epsilon}$ in the sense that
$\left|\frac{\tilde{\epsilon}}{\epsilon}-1\right|\leq \eta$
if $P_R\leq\phi^{-1}\left(\frac{\eta}{1+\eta}\right)$ for some small error tolerance $\eta>0$. In other words, $\phi(P_R)$ can be used to compute the range of envelopes over which the relative tail approximation error is less than $\eta$.

Table~\ref{tab:tail_approx} summarizes the tail approximations that are derived in the sequel. 
%\PP{The bracketed entry $()^*$ is very confusing, better to explain verbally instead of usign this unnecessary complication. For example, in order to understand the bracketed entry $(\tfrac{1}{2},\tfrac{3}{4})^*$, I need to read the text, but then after reading the text, why do I need the table? The idea of the table is to be self-sufficient quick reference.}\PE{well can remove the bracketed entries all together, as is a TW 3W thing..and those are the least important distributions at all}

\subsection{Two-Wave Model (TW)}
\label{sec:TW}

We start with the common Two-Wave channel model \cite{Durgin2002}, where $\mu=1$, $N=2$ and $L=0$, i.e., single cluster, two specular and no diffuse components.
The envelope PDF is given by:
\begin{equation}
f_{\mathrm{TW}}(r)=\frac{2r}{\pi A_{\mathrm{TW}}\sqrt{\Delta^2-\left(1-\frac{r^2}{A_{\mathrm{TW}}}\right)^2}}, 
\end{equation}
where $A_{\mathrm{TW}}=\rho_1^2+\rho_2^2$, $\Delta$ is given by (\ref{DeltaDefinition}) and $r \in [r_{\min},r_{\max}]=\left[\sqrt{A_{\mathrm{TW}}(1-\Delta)}, \sqrt{A_{\mathrm{TW}}(1+\Delta)}\right]$.
By putting $f_{\mathrm{TW}}(r)$ in (\ref{eq:EpsilonIntegral}) we obtain the CDF:
\begin{equation}\label{eq:two_wave_cdf}
\epsilon=F_{\mathrm{TW}}(P_R)=\frac{1}{2}-\frac{1}{\pi} \mathrm{asin} \left( \frac{1-\frac{P_R}{A_{\mathrm{TW}}}}{\Delta}\right).
\end{equation}
%which results in 
%\begin{equation} \label{eq:PR_twowave}
%P_R=F^{-1}_{\mathrm{TW}}(\epsilon)=A_{\mathrm{TW}}(1-\Delta \cos(\pi \epsilon)) %\approx A_{\mathrm{TW}} \left(1-\Delta + \Delta\frac{(\pi \epsilon)^2}{2}\right)
%\end{equation}
{
%We first treat the special case $\Delta=1$; 
Bounding $\epsilon$ from below leads to the tail approximation:
\begin{equation} \label{eq:PR_twowave}
\tilde{\epsilon} = \frac{1}{\pi}\sqrt{\frac{2}{A_{\mathrm{TW}}}}\sqrt{P_R^{\star}},
%P_R=F^{-1}_{\mathrm{TW}}(\epsilon)=A_{\mathrm{TW}}(1-\Delta \cos(\pi \epsilon)) \approx A_{\mathrm{TW}} \left(1-\Delta + \Delta\frac{(\pi \epsilon)^2}{2}\right)
\end{equation}
where $P_R^{\star} = \frac{1}{\Delta}P_R - \frac{1-\Delta}{\Delta}A_{\mathrm{TW}}$.
The approximation error function is
(see Appendix \ref{app:App Error functions}):
\begin{equation}
\phi(P_R) = \frac{4}{3}\sqrt{\frac{A_{\mathrm{TW}}}{2}}\frac{(A_{\mathrm{TW}}+P_R^{\star})P_R^{\star}}{\sqrt{(2A_{\mathrm{TW}}-P_R^{\star})^3}}.
\end{equation}
The upper bound on the power for error tolerance $\eta$ %$\varepsilon
can be evaluated numerically.
%It should be noted that the power law approximation \eqref{eq:PR_twowave} is valid for $P_R\geq A_{\mathrm{TW}}(1-\Delta)$.
%It should be noted that the tail approximation $\tilde{\epsilon}$ is actually a lower bound on $\epsilon$.
%\begin{equation}\label{eq:two_app_err_fun}
%\phi(P_R) = \frac{4}{3}\sqrt{\frac{A_{\mathrm{TW}}}{2}}\frac{(A_{\mathrm{TW}}+P_R)P_R}{\sqrt{(2A_{\mathrm{TW}}-P_R)^3}}.
%\end{equation}
%When $\Delta<1$, $P_R\geq A_{\mathrm{TW}}(1-\Delta)>0$.
}

\begin{figure}
\centering
\includegraphics[scale=0.6]{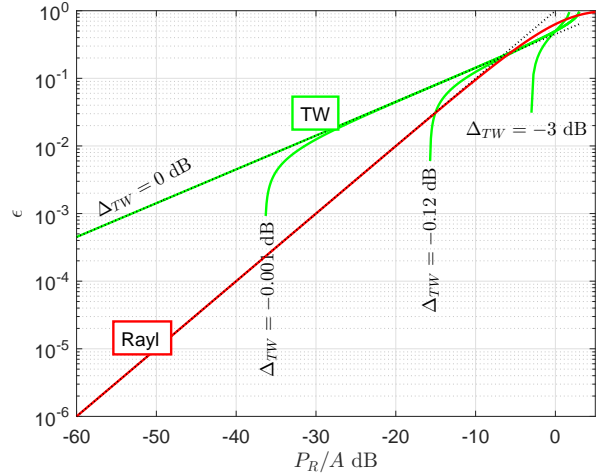} 
\caption{Two-wave \eqref{eq:two_wave_cdf} and the classical Rayleigh \eqref{eq:CDF_Rayleigh} CDFs and their tail approximations \eqref{eq:PR_twowave}, \eqref{eq:PR_rayleigh} (black dotted lines).
%\PE{well the TW approx is not in .. should I put it??}): \MA{@PATRICK you can add additional simulation parameters here}.
For Two-wave: $\rho_1=1$, while $\rho_2$ is set to correspond to the $\Delta$ \eqref{DeltaDefinition} given at each curve.}
\label{fig:Fig_2W_3W}
\end{figure} 
%Furthermore, the logarithmic relation for $\Delta=1$ is:
%\begin{equation}
%\log \epsilon= \mathrm{const.}+\frac{1}{2} \log P_R
%\end{equation}
%i.e. with a log-log linear slope of $\beta\approx \frac{1}{2}$.
% \textbf{I still like the Table much better for hands on information than having to plow through pages of equations. Table is very useful for an applications engineer-> will be supporting the Diversity section directly. Now paper is turning more away from that and into a 'dry' theory exercise.}
{
Fig.~\ref{fig:Fig_2W_3W} depicts the tail $\epsilon$ for the TW channel. %\PP{This is a very dense figure depicting also 3W channel, before even it has been discussed. This does not correspond to the way we are presenting the channels. Hence, Figure 1 should compare 2W and Rayleigh, Figure 2 should compare Weibull, Rice and Nakagami.}. \PE{Well TW and 3W support each other graphically.. if you remove 3W i suggest remove TW as well - and for that matter Rayleigh can just be incoorporated in any of the other figs.} %  CDF of $\frac{P_R}{A_{\mathrm{TW}}}$.
When $\Delta<1$, the tail falls abruptly to zero  at $P_R=A_{\mathrm{TW}}(1-\Delta)$.
However, as it is seen from Fig.~\ref{fig:Fig_2W_3W}, the log-log slope that precedes this abrupt transition to zero is $\frac{1}{2}$ (half a decade per $10$ dB), which can be also see from \eqref{eq:PR_twowave}.
In the singular case $\Delta=1$ ($0$ dB),
the tail approximation is given by \eqref{eq:PR_twowave} with $P_R^{\star} = P_R$ and the slope continues until $-\infty$ dB\footnote{The case $\Delta \approx 1$ has been referred to in the literature as hyper-Rayleigh fading \cite{Frolik}}.
For example, if the log-log slope of $\frac{1}{2}$ should be present at $\epsilon=10^{-6}$, then we need to have $\Delta> -4\cdot10^{-6}\mathrm{dB}$, i.e. $\rho_2$ very close to $\rho_1$, which is unlikely in practice due to the losses of the reflected wave.
Hence, the two-wave model should be used with high caution when evaluating URLLC scenarios.
}

\subsection{Rayleigh Channel (Rayl)}\label{sec:Rayl}

This model, adopted in many wireless studies, has $\mu=1$, $N=0$ and $L=1$ (single cluster and diffuse component and no specular components) and the envelope PDF is \cite{RVJBA}:
\begin{equation}\label{eq:p_rayleigh}
f_{\mathrm{Rayl}}(r)=\dfrac{2r}{A_{\mathrm{Rayl}}}e^{-\frac{r^{2}}{A_{\mathrm{Rayl}}}},
\end{equation}
with average power $A_{\mathrm{Rayl}}=2\sigma^2$.
The CDF follows readily as:
\begin{equation}\label{eq:CDF_Rayleigh}
\epsilon=F_{\mathrm{Rayl}}(P_R)=1-e^{-\frac{P_R}{A_{\mathrm{Rayl}}}},
\end{equation}
{which can be upper bounded by retaining only the first term in the Taylor expansion, resulting in the following simple power law approximation:
%which for $\epsilon \rightarrow 0$ is approximated by
\begin{equation} \label{eq:PR_rayleigh}
\tilde{\epsilon} = \frac{P_R}{A_{\mathrm{Rayl}}},
%P_R=F^{-1}_{\mathrm{Rayl}}(\epsilon)=-A_{\mathrm{Rayl}}\ln(1-\epsilon) \approx A_{\mathrm{Rayl}} \epsilon,  
\end{equation}
also known as the Rayleigh rule of thumb ``10dB outage margin per decade probability'' due to a log-log slope of $\beta = 1$. 
The approximation error function can be derived via an upper bound on the Taylor remainder, yielding the simple form (see Appendix A):
\begin{equation}\label{eq:error_func}
\phi(P_R) = \frac{P_R}{2A_{\mathrm{Rayl}}}.
\end{equation}

\subsection{Rician Channel (Rice)}\label{sec:Rice}

This is an extension of the Rayleigh channel, featuring a specular component in addition to the diffuse one. 
The average received power is \(A_{\mathrm{Rice}}=\rho_1^2+2\sigma^{2} =2\sigma^{2}(k_1+1)\), where $k_1=\frac{\rho_1^2}{2\sigma^{2}}$ is the Rician \emph{k-factor} and the PDF of the received envelope is~\cite{RVJBA}:
\begin{equation}\label{eq:p_rice}
f_{\mathrm{Rice}}(r)=f_{\mathrm{Rayl}}(r)e^{-k_1}I_{0}\left(\frac{r}{\sigma}\sqrt{2k_{1}}\right)
\end{equation}
where $I_0(\cdot)$ is the modified Bessel function of 1$^\text{st}$ kind and 0$^\text{th}$ order.
{The tail can then be expressed in closed form in terms of the 1$^\text{st}$ order Marcum Q-function as follows:
\begin{equation}\label{eq:cdf_rice}
\epsilon = F_{\mathrm{Rice}}(P_R) = 1 - Q_1\left(\sqrt{2k_1},\sqrt{2\frac{P_R}{A_{\mathrm{Rice}}}(k_1+1)}\right).
\end{equation}
Bounding $\epsilon$ from below via 1$^\text{st}$ polynomial expansion of $Q_1$, we arrive at the tail approximation:
\begin{equation}\label{eq:lowtail_rice}
\tilde{\epsilon}=\frac{P_R}{A_{\mathrm{Rice}}}(k_1 + 1)e^{-k_1}.
\end{equation}
%Note that the above approximation is a tight lower bound which can be also derived from \eqref{eq:p_rice} by bounding $I_0(\cdot)\geq 1$, valid for small values of the argument.
The approximation error function obtains the form (see Appendix \ref{app:App Error functions}):
\begin{equation}\label{eq:upp_err_func_rice}
\phi(P_R) = e^{\frac{k_1}{2}}\left(e^{\frac{P_R}{A_{\mathrm{Rice}}}(k_1 + 1)} - 1\right),
\end{equation}
The tail approximation of the Rician channel in URLLC-relevant regime has Rayleigh slope $\beta = 1$. However, Fig.~\ref{fig:Fig_TWDP_Rice} shows that before attaining the slope $\beta=1$, the Rician CDF has a steeper slope compared to the Rayleigh one. In the context of wireless communications, this can be interpreted as an increased diversity order offered by the Ricean distribution. The lower the $k_1-$factor, the sooner the slope
becomes identical to the Rayleigh one; in other words, as $k_1$ increases, $P_R$ decreases for fixed error tolerance $\eta$ which can be also see from \eqref{eq:upp_err_func_rice}. %The change of the slope in Fig.~\ref{fig:Fig_TWDP_Rice3a} is indicated by circles. 
}
%The CDF lower tail can be extracted from \cite{Pent} polynomial expansion of the Marcum Q function, while here we derive lower and upper bounds on the CDF.

%To get the lower bound, we observe that $I_0(x)\geq 1$ and $\lim_{x\rightarrow 0}I_0(x)=1$. This corresponds to $\frac{P_R}{\bar {P}_{\mathrm{Rice}}} \ll  \tfrac{1}{4k_1(k_1+1)}$ and the tail of the CDF  translates to a scaled Rayleigh as: 
%\begin{equation}\label{eq:e_rice}
%\epsilon=F_{\mathrm{Rice}}(P_R) \geq F_{\mathrm{Rayl}}\left(r\right)e^{-k_1} =\left(1-e^{-\frac{P_R}{A_{\mathrm{Rice}}}(k_1+1)}\right)e^{-k_1}
%\end{equation}

%An upper bound can be derived by using the result from \cite{Baricz} that 
%$I_0(x)< e^{x^2/4}$. With this we can approximate $f_{Rice}\left(r\right)\approx\frac{r}{\sigma^{2}}e^{-\frac{r^{2}}{2\sigma^{2}}\left(1-k_1\right)}e^{-k_1}$
% , with relative error $\left|\epsilon_{\mathrm{rel}} \left(x\right)\right|=\left|\frac{I_{0a}\left(x\right)}{I_{0}\left(x\right)}-1\right| < 2\cdot10^{-2}$ for $x=1$ and decaying fast by order $x^4$ as $x\rightarrow0$. 
%and the CDF tail is:
%\begin{eqnarray}\label{eq:e_rice2}
%\epsilon &=& F_{\mathrm{Rice}}(P_R)< \left(1-e^{-\frac{P_R}{A_{\mathrm{Rice}}}\left(1-k_1^2\right)}\right) \frac{e^{-k_1}}{1-k_1} 
%&\approx& \frac{P_R}{A}(1-k_1^2)\frac{e^{-k_1}}{1-k_1} =\frac{P_R}{A}(1+k_1)e^{-k_1}
%\end{eqnarray}

%Using same approximation as in \eqref{eq:log_e_rice}, we can invert \eqref{eq:e_rice} as
%\begin{equation}
%P_R = F^{-1}_{\mathrm{Rice}} (\epsilon) \approx \epsilon A_{\mathrm{Rice}} \frac{e^{k_1}}{k_1+1}
%\end{equation}

\begin{figure}
\centering
\includegraphics[scale=0.6]{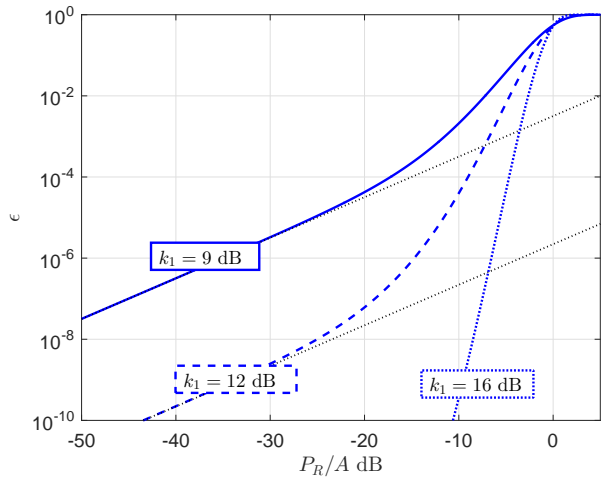} 
\caption{Rician CDF \eqref{eq:cdf_rice} and its tail approximation \eqref{eq:lowtail_rice} (black dotted lines): % \MA{All additional params here.}
The Rician \emph{k-factor} $k_1$ is indicated at the respective curves.}
%SOME PLACES WE HAVE SEVERAL TAIL EXPRES. = PREFER HAVING SPECIFC EQ NUMB. in }}
\label{fig:Fig_TWDP_Rice}
\end{figure}

\subsection{Weibull Channel (Wei)}\label{sec:Wei}
The Weibull channel is a generalization of the Rayleigh model, where the diffuse component is given by $|V_{\mathrm{dif}}|=\sqrt{(X_R^2+X_I^2)^{1/\beta}}$ with $\beta \neq 1$. This model has been used in empirical studies as it offers increased freedom to fit the modeling of the diffuse part \cite{Lorenz1979,hashemi1993}.
As in the Rayleigh case, here we also have only a diffuse component, but the received envelope follows Weibull distribution \cite{bessate2016}:
%,adawi1988,
{
%The received envelope 
%$r=|V_{\mathrm{dif}}|$ has a PDF given by :
\begin{equation}\label{eq:f_wei}
f _{\mathrm{Wei}}(r)=\frac{2\beta r^{2\beta-1}}{2\sigma^2 } e^{-\frac{r^{2\beta}}{2\sigma^2 }}
\end{equation}
with
%shape parameter  \(k=2\beta>0\)  and scale parameter \(\lambda=\left(2\sigma^2 \right)^{1/2\beta}>0\),
\(A_{\mathrm{Wei}}=\left(2\sigma^{2}\right)^{1/\beta} \Gamma(1+1/\beta)\).
For \(\beta=\tfrac{1}{2}\) we get the TW case, while \(\beta=1\) leads to the Rayleigh case.
The tail is given as: 
\begin{equation} \label{eq:e_wei}
\epsilon=F_{\mathrm{Wei}}(P_R)=1-e^{-\left(\Gamma(1+1/\beta)\frac{P_R}{A_{Wei}}\right)^{\beta}}.
\end{equation}
%\begin{equation}\label{eq:PR_wei}
%P_R=F^{-1}_{\mathrm{Wei}}(\epsilon)=  \frac{A_{\mathrm{Wei}} }{\Gamma(1+1/\beta)} \left(-\ln(1-\epsilon)\right)^{1/\beta},
%\end{equation}
Using first order Taylor expansion, we obtain the following tail approximation:
\begin{equation}\label{eq:tail_app_wei}
\tilde{\epsilon} = \left( \Gamma(1 + 1/\beta)\frac{P_R}{A_{\mathrm{Wei}}} \right)^{\beta}.
\end{equation}
%\begin{equation}\label{eq:log_e_wei}
%\log \epsilon\approx  \beta\left(\log \frac{P_R}{A_{\mathrm{Wei}}}+\log\Gamma(1+1/\beta)\right)
%\end{equation}
%where we have used the same type of approximation as in the Rayleigh case. 
Here $\beta$ denotes the log-log slope {and an example with $\beta=2$ is shown in Fig. \ref{fig:Fig_Wei_Nak_kappamu}}. 
% The narrow %'sharp'
% shoulder means that the approximation in (\ref{eq:tail_app_wei}) is accurate close up to the mean level. % \PP{If you want to illustrate Weibull, then compare to Rayleigh, Rice, and Nakagami, and the figure should be immediately in this section, not after the figure in which you are comparing Rice and TWDP.} \PE{OK can do that BUt TWDP and Rice also complemnet each other.. so then suggest remove illustration of TWDP and it's discusson wrt teh Fig}
The approximation error function for the Weibull channel is given by (see Appendix \ref{app:App Error functions}):
\begin{equation}\label{eq:phi_wei}
\phi(P_R) = \frac{\left( \Gamma(1+ 1/\beta)\frac{P_R}{A_{\mathrm{Wei}}} \right)^{\beta}}{1+ \left( \Gamma(1+ 1/\beta)\frac{P_R}{A_{\mathrm{Wei}}} \right)^{\beta}}.
\end{equation}

\subsection{Nakagami-m Channel (Nak)}\label{sec:Nak}
The envelope of this model behaves similarly to the Weibull model, although the diffuse component is modeled differently as $r=|V_{\mathrm{dif}}|=\sqrt{\sum_{i=1}^m (X_{Ri}^2+X_{Ii}^2)}$, with $m$ integer. This model can be interpreted as an incoherent sum of $m$ i.i.d. Rayleigh-type clusters, each with mean power $2\sigma^2$ and total power \(A_{\mathrm{Nak}}=m\cdot2\sigma^2\).  
The PDF of the envelope $r$ is given by \cite{Nakagami1958}:
\begin{equation}\label{eq:f_nak}
f _{\mathrm{Nak}}(r)=\frac{2m^m}{\Gamma(m)A_{\mathrm{Nak}}^m} r^{2m-1} e^{-m\frac{r^2}{A_{\mathrm{Nak}}}}=\frac{2}{r\Gamma(m)}\left(\frac{r^2}{2\sigma^2}\right)^m  e^{-\frac{r^2}{2\sigma^2 }},
\end{equation}
where we interpret $m \in \R \geq \tfrac{1}{2}$ for generality.
For $m=\tfrac{1}{2}$ we get an exponential, while for $m=1$ a Rayleigh distribution.
The CDF is given as:
\begin{equation}\label{eq:e_nak}
\epsilon=F_{\mathrm{Nak}}(P_R)=\frac{\gamma \left(m;m\frac{P_R}{A_{Nak}}\right)}{\Gamma(m)}
\end{equation}
where $\gamma(a,x)$ is the lower incomplete gamma function.
The power law tail approximation can be obtained via the upper bound $\gamma(a;x)\leq x^a/a$ \cite{Jameson2016}, resulting in:
%\begin{align}\label{eq:e_bounds_nak}
% \tilde{\epsilon} = \frac{m^{m}}{\Gamma(m+1)} \left(\frac{P_R}{A_{Nak}}\right)^m
%\end{align}
%The power law approximation of the tail is given by:
\begin{equation}\label{eq:e_approx_nak}
\tilde{\epsilon} = \frac{m^{m}}{\Gamma(m+1)} \left(\frac{P_R}{A_{Nak}}\right)^m.
\end{equation}
We see that \eqref{eq:e_approx_nak} has the same flexibility and slope behavior %wrt \PP{Please use whole words, not short versions like 'wrt' or 'approx.'} slope
as the Weibull model (for $m\geq\tfrac{1}{2}$), but a different offset. This can be also observed in Fig. \ref{fig:Fig_Wei_Nak_kappamu} with $m=2$, where the wide shoulder sends the tail (\ref{eq:e_approx_nak}) to lower levels compared with the Weibull case.
By bounding the lower incomplete gamma function in \eqref{eq:e_nak} from both sides, i.e., $e^{-x}x^a/a\leq\gamma(a;x)\leq x^a/a$ \cite{Jameson2016}, we derive the approximation error function:
%\begin{equation}\label{eq:low_upp_tail_nak}
%(1-\phi_{\mathrm{Nak}}(P_R))\tilde{\epsilon} \leq \epsilon \leq \tilde{\epsilon},
%\end{equation}
%where $\phi_{\mathrm{Nak}}(P_R)$ is simply:
\begin{equation}\label{eq:err_func_nak}
\phi(P_R) = 1 - e^{-m\frac{P_R}{A_{\mathrm{Nak}}}}\leq  e^{m\frac{P_R}{A_{\mathrm{Nak}}}} - 1.
\end{equation}

\subsection{$\kappa-\mu$ Channel ($\kappa\mu$)}\label{sec:km}
The $\kappa-\mu$ model was developed in \cite{Yacoub2007} as a generalization to the Nakagami model, by considering incoherent sum of $\mu$ Rician type clusters, i.e.  envelope $r=\sqrt{\sum_{i=1}^\mu (X_{Ri}+p_i)^2+(X_{Ii}+q_i)^2}$ where $X_{Ri}+jX_{Ii}$ are complex Gaussian diffuse components (all same mean power $2\sigma^2$) and $p_i+jq_i$ the corresponding specular components with arbitrary power $\rho_i^2=p_i^2+q_i^2$. Here $\kappa$ is a generalized Rician type \emph{k-factor} defined in (\ref{eq:k-factor}).
%$\kappa=\frac{\sum_{i=1}^{\mu}\rho_i^2}{\mu \cdot 2\sigma^2}$ [THIS DEFINITION MOVED UP FRONT IN SYSTEM SECTION]
%is a Rician type k-factor, i.e. ratio of all dominant component power to all diffuse power. 
Consequently, the total mean power is \(A_{\mathrm{\kappa\mu}}=\mu (1+\kappa)\cdot2\sigma^2\) and the PDF of $r$ is given by \cite{Yacoub2007}:
\begin{equation}\label{eq:f_mukappa}
f _{\mathrm{\kappa\mu}}(r)=\frac{2(\kappa\mu)^{(1-\mu)/2}}{e^{\kappa\mu}\sqrt{2\sigma^2}} \left( \frac{r}{\sqrt{2\sigma^2}}\right)^{\mu} e^{\left(-\frac{r^2}{2\sigma^2}\right)}  I_{\mu-1}\left[2\sqrt{\kappa\mu\frac{r^2}{2\sigma^2}} \right], \mu\geq 0
\end{equation}
Again, for generality, we interpret $\mu \in \R$. The CDF in closed form is described via the generalized Marcum Q function, from \cite{Yacoub2007}:
\begin{equation}\label{eq:cdf_kappamu}
F_{\kappa\mu}(P_R)=1-Q_{\mu}\left(\sqrt{2\kappa\mu},\sqrt{2(1+\kappa)\mu P_R/A_{\kappa\mu}}\right)
\end{equation}
%Tight bounds of $Q_{\mu}(a,b)$ are given in \cite[(66),(69)]{Bariz2009}, while polynomial expansion for fractional order $\mu$ is found in \cite[(9),(14)]{Andras2011}. 
Using a first-order polynomial expansion of the generalized Marcum Q-function, we obtain the following tail approximation: % (with $\gamma$ approx. in \cite{Jameson2016}): 
\begin{equation}\label{eq:e_kappamu}
\tilde{\epsilon} = \frac{\left(e^{-\kappa}(\kappa+1)\mu \right)^{\mu}}{\Gamma(\mu+1)} \left(\frac{P_R}{A_{\mathrm{\kappa\mu}}}\right)^{\mu}=F_{Nak}\left(\tfrac{P_R}{A_{\mathrm{\kappa\mu}}};\mu\right) F_{Rice}(1;\kappa)^\mu %= \tilde{\epsilon}
\end{equation}
i.e. a multiplicative form of the previous Rician and Nakagami-m tail approximation.
%This leads to the inversion:  
%\begin{equation}\label{eq:PR_kappamu}
%P_R=F^{-1}_{\mathrm{\kappa\mu}}(\tilde{\epsilon}) = \frac{A_{\mathrm{\kappa\mu}}}{e^{-\kappa}(\kappa+1)\mu}\Gamma(\mu+1)^{1/\mu}\epsilon^{1/\mu}
%\end{equation}
%\begin{equation}\label{eq:log_e_kappamu}
%\log \epsilon\approx  \mu\left(\log \frac{P_R}{A_{\mathrm{\kappa\mu}}}+\log\left(e^{-\kappa}(\kappa+1)\mu \right)\right)-\Gamma(1+\mu)
%\end{equation}
The approximation error function obtains the simple form (see Appendix \ref{app:App Error functions}):
\begin{equation}\label{eq:up_bound_kappamu}
\phi(P_R) = e^{\frac{\kappa\mu}{2}}\left(e^{(\kappa+1)\mu\frac{P_R}{A_{\mathrm{\kappa\mu}}}}-1\right).
\end{equation}
%from which we derive the upper bound on $P_R$ for given error tolerance $\varepsilon$:
%\begin{equation}\label{eq:upp_PR_kappamu}
%P_R \leq \frac{A_{\mathrm{\kappa\mu}}}{(k+1)\mu}\ln(1 + \varepsilon e^{-\frac{\kappa\mu}{2}}).
%\end{equation}
We see that both $\tilde{\epsilon}$ and $\phi(P_R)$ reduce to the forms derived earlier as special cases; specifically, for $\mu=1$ the $\kappa-\mu$ model reduces to a Rician situation, while for $\kappa=0$ the Nakagami-m situation emerges.

The tail (\ref{eq:e_kappamu}) in a typical Rician  setting ($\kappa=3.9,\mu=2$) is seen in Fig. \ref{fig:Fig_Wei_Nak_kappamu}, where it can be seen that the tail is pushed to lower probabilities compared to Nakagami and Weibull models.

\begin{figure}
\centering
\includegraphics[scale=0.6]{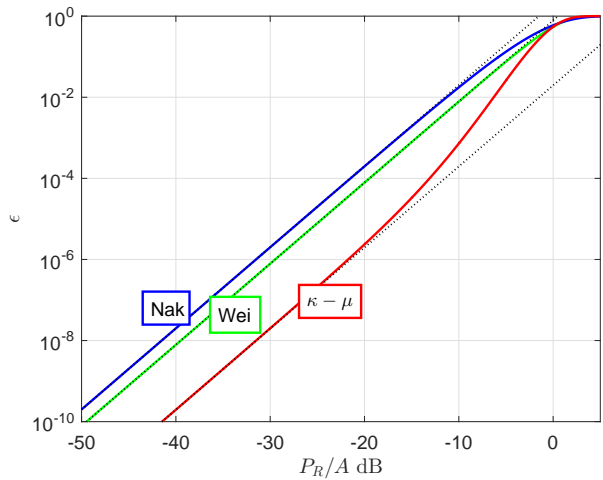} 
\caption{Nakagami-m \eqref{eq:e_nak} ($m=2$), Weibull \eqref{eq:e_wei} ($\beta=2$) and $\kappa-\mu$ \eqref{eq:cdf_kappamu} ($\kappa=3.9, \mu=2$) CDFs and their tail approximations \eqref{eq:e_approx_nak},\eqref{eq:tail_app_wei},\eqref{eq:e_kappamu} (black dotted lines). %(\MA{@PATRICK all other parameters here}). %and TWDP (num. integration of (\ref{eq:f_TWDP}) with 256 digits prec. and N=1000 terms) distributions for different values of the relevant parameters. 
 %(\ref{eq:e_approx_twdp}). Solid dots: 1dB deviation tail approx. vs. distributions
}
\label{fig:Fig_Wei_Nak_kappamu}
\end{figure}

\subsection{Generalizations
\label{sec:generalizations}}
We analyze several generalizations of the channels presented in the previous subsection and derive their power law tail approximations.
First, we explore the transition of the behavior from few paths to many paths. In this sense, we expand the previous TW model to cater for $3-$vector components or include a diffuse part that is generalized compared to the models with diffuse part in the previous section. As it will be shown, both cases result in a tail behavior that conforms to a behavior dominated by diffuse components. In other words, three specular components can be sufficient to produce the behavior of a Rayleigh diffuse component at URLLC levels.

Another important generalization is to use combined short and long term processes, particularly when such are inseparable. We consider three combined models in following sub-sections: 1) Log-normal shadowed Rayleigh fading, i.e. the Suzuki distribution typically used to model Macro-cell behavior \cite{RVJBA}; 2) $\kappa-\mu$ fading with Nakagami-m shadowing; and 3) $\kappa-\mu$ fading with inverse Gamma distributed shadowing. While 1) is a classical case, 2) and 3) have been found useful for modeling close range propagation \cite{Paris2014,Yoo2016,Yoo2015}.

We can think of the combined channel models to be applicable to the following situation. When there is only short term block fading, the outage probability can be controlled by selecting the rate $R$ according to the known average power of the short term channel. Equally important is the specular component balancing or ratio towards the diffuse parts, captured by $\Delta$, $\kappa$ in (\ref{DeltaDefinition}), (\ref{eq:k-factor}). However, when the sender does not have a reliable estimate of the average power (or impact of specular  components), then this uncertainty can be modeled by assuming that the average power or $\Delta$, $\kappa$ are random variables.
The independent sampling from the shadowing distribution is a pessimistic case that assumes sporadic transmissions, sufficiently separated in time.  

%\textbf{\PE{[MOve here and rephrase Petars section on shadowing currently in the SUzuki section]}\MA{Maybe put it last, since it is a bit exotic.}} \textbf{\PE{[wrt Suzuki = well one of most referred to combined distr, though not used so much in praxis..so not so much exotic as classic..BUT Did not think to move WHOLE SUzuki section up front .. just steal the paragraphs discussing inseparability of effects = reason to model combined distr. in general]}}
\subsubsection{Three-Wave Model (3W)}
We consider the Three-Wave generalization of the TW model.
Here $N=3$, $V_{\DIF}=0$, received envelope $r=|\rho_1+\rho_2e^{j\phi_2}+\rho_3e^{j\phi_3}|
$ and average power $A_{\mathrm{3W}} = \sum_{n=1}^3{\rho_n^2}$.
The probability density function \cite{Durgin2002}
%,Nicholson}
is given by:
\begin{equation} \label{eq:3Wdistribution}
f_{\mathrm{3W}}(r)=\left\{ \begin{array}{ll}
\frac{\sqrt{r}}{\pi^2 \sqrt{\rho_1\rho_2\rho_3}} K \left( \frac{\Delta_r^2}{\rho_1\rho_2\rho_3 r}\right) & \Delta_r^2 \leq\rho_1\rho_2\rho_3 r \\
\frac{r}{\pi^2 \Delta_r} K \left( \frac{\rho_1\rho_2\rho_3 r}{\Delta_r^2}\right) & \Delta_r^2 > \rho_1\rho_2\rho_3 r
\end{array} \right.
\end{equation}
for $r \in [r_{\min},r_{\max}]$, and it is $0$ otherwise, with $r_{\min}=\max(2 \max(\rho_1,\rho_2,\rho_3)-\rho_1-\rho_2-\rho_3,0),~r_{\max}= \rho_1+\rho_2+\rho_3$.
In (\ref{eq:3Wdistribution}), $K(\cdot)$ is an elliptic integral of the first kind\footnote{Convention of \cite{Durgin2002}
%Nicholson,
 is $K(m)$ with argument $m=k^2$ (instead of  $K(k)$ with modulus $k$).}  and 
the quantity $\Delta_r$ is defined as:
\begin{equation}\label{eq:Deltar}
\Delta_r^2=\frac{1}{16} [(r+\rho_1)^2-(\rho_2-\rho_3)^2] [(\rho_2+\rho_3)^2-(r-\rho_1)^2].
\end{equation}
Without losing generality, we can take $\rho_1\geq\rho_2\geq\rho_3$  and define the difference $\Delta_{\rho}=\rho_1 - (\rho_2+\rho_3)$, such that  $r_{\min}=\max(\Delta_{\rho},0)$. Three cases can be considered: (1) $r_{\min}=0$ when $\Delta_{\rho}<0$; (2) $r_{\min}=0$ and $\Delta_{\rho}=0$; and (3)  
$r_{\min}>0$ otherwise. Here we treat the case $\Delta_{\rho}<0$, which sets the basis for the reader to treat the other two cases. 
The integral (\ref{eq:EpsilonIntegral}) is evaluated for values $r \in [0,\sqrt{P_R}]$ that are very small and taking $r \rightarrow 0$:
\begin{eqnarray}
\lim_{r \rightarrow 0}\Delta_r^2 = [\rho_1^2-(\rho_2-\rho_3)^2][(\rho_2+\rho_3)^2-\rho_1^2] >0
\end{eqnarray}
which implies that $\Delta_r^2 > \rho_1\rho_2\rho_3 r$ holds in (\ref{eq:3Wdistribution}).
With $r \rightarrow 0$:
\begin{equation}\label{eq:3W_approx}
f_{\mathrm{3W}}(r) \stackrel{r \rightarrow 0}{\rightarrow} \frac{r}{\pi^2 \Delta_r} K \left( \frac{\rho_1\rho_2\rho_3 r}{\Delta_r^2}\right) \approx \frac{r}{\pi^2 \Delta_r} \frac{\pi}{2}
\end{equation}
where we have used $\lim_{x \rightarrow 0} K(x)=\frac{\pi}{2}$.
Approximating $\Delta_r^2$ as a constant for small values of $r$, we get the following tail approximation:
\begin{equation}\label{eq:e_3W}
\tilde{\epsilon}=\frac{r^2}{4\pi\Delta_r}=\frac{P_R}{4\pi\Delta_r}
\end{equation}
such that the log-log linear slope is $\beta \approx 1$. In the singular case $r_{\min}=0$ and $\Delta_{\rho}=0$ it can be shown that $\beta=\tfrac{3}{4}$, while the case $r_{\min}>0$ has a slope of $\tfrac{1}{2}, \tfrac{3}{4}$ or $1$, before an abrupt fall to zero when $P_R=r_{\min}^2$. 

\begin{figure}
\centering
\includegraphics[scale=0.6]{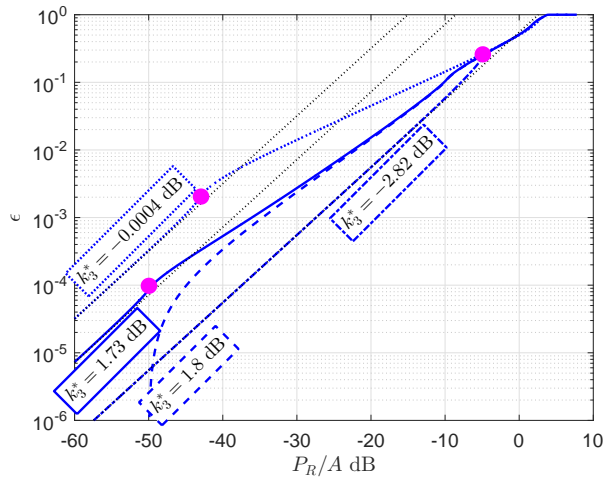} 
\caption{Three Wave CDF (numerical integration of \eqref{eq:3Wdistribution}) and the tail approximation \eqref{eq:e_3W} (black dotted line):
%\MA{@PATRICK all necessary descriptive params here.}
Here $\rho_1=1$ while $\rho_2$ and $\rho_3$ are set according to $k_3^*$ indicated at the curves. The solid dots depicts the locations of the ``breaking points'' where the asymptotic slopes change (as determined by the value of $|\Delta_{\rho}|^2$). % \PE{levels}. \PP{I did not get this with the solid dots.}\PE{This is 'break point' values where transition from one slope to the other happen/ or hitting rmin=max(Delta rho,0)}
%\PE{Suggest to follow the sentence to end w something like: ' acting as break point between different asymptote slopes'}
}
\label{fig:Fig_3W}
\end{figure}

The 3W CDF %generated with high-precision numerical integration, 
is shown in Fig. \ref{fig:Fig_3W} for $\rho_1=1$ and different variations of $\rho_2$ and $\rho_3$. The curves are labeled by $k_3^*=\frac{\rho_1^2}{(\rho_2^2+\rho_3^2)}$ \footnote{Note that this metric differs from the \emph{k-factor} which involves diffuse parts. $k_3^*$ is the ratio between specular component powers only}. 
We select to represent two cases with identical $10\log\tfrac{|\Delta_{\rho}|^2}{A_{\mathrm{3W}}}= -50$dB that are seen to diverge significantly when $\frac{P_R}{A_{\mathrm{3W}}}<-40$dB. The difference between these two cases emerges due to the different sign of $\Delta_\rho$% \PE{(0.0041 vs -0.0041)}
, which in one case results in $r_{\min}>0$ ($\rho_2$=0.7850, $\rho_3$=0.2109) and in the other case $r_{\min}=0$ ($\rho_2$=0.7914, $\rho_3$=0.2126). The latter case has a log-log slope of $\beta=1$, identical to the Rayleigh distribution. Hence, if the sum of the two smallest
%other specular
components 
%is large enough
%sufficiently strong to
can cancel and overshoot the strongest component,
%(i.e. $\Delta_\rho<0$)
the URLLC-level behavior of the 3W model is practically identical to that of a Rayleigh channel in terms of a
slope. %\PP{We need a new figure in which we put \textbf{only two selected 3W curves}, not as many as you have on Figure 1.}
%, with a certain offset from the actual Rayleigh curve.}
%In other words, three specular components that lead to $\Delta_\rho<0$ are sufficient to produce the behavior of a Rayleigh diffuse component.}  
%\PE{Or remove any TW, 3W TWDP Fig all together}

\subsubsection{Two-Wave Diffuse Power (TWDP) Channel}
In this model $N=2$ and $V_{\DIF}$, with envelope $r=|\rho_1+\rho_2+V_{\DIF}|$ and average received power \(A_{\mathrm{TWDP}}=\rho_1^2+\rho_2^2+2\sigma^{2} \) \cite{Durgin2002}.
The PDF is obtained by averaging of the Rician PDF \cite{rao2015mgf}:
\begin{equation}\label{eq:f_TWDP}
f_{\mathrm{TWDP}}(r)=\frac{1}{2\pi}\int_{0}^{2\pi}f_{\mathrm{Rice}}\left(r;k_{2}\left[1+\Delta\cos\left({\psi}\right)\right]\right)d{\psi}
\end{equation}
with $\Delta$ defined in (\ref{DeltaDefinition})
and $k_2$ in (\ref{eq:k-factor}).
The integration over $\psi$ involves only $I_0(\cdot)$ and the exponential terms in (\ref{eq:p_rice}). 
Using $I_0 \geq  1$ for $\frac{P_R}{A_{\mathrm{TWDP}}} \ll  \tfrac{1}{4k_2(k_2+1)}$, this integration is $\sim\frac{1}{2\pi}\int_{0}^{2\pi}e^{k_{2}\Delta\cos\psi}\cdot1\,d\psi=I_{0}\left(k_{2}\Delta\right)$, i.e. it leads to a constant with respect to $r$. Hence, the tail can be lower-bounded through a scaled Rician tail:
\begin{eqnarray}\label{eq:e_approx_twdp}
\epsilon = F_{\mathrm{TWDP}}(P_R) \geq  F_{\mathrm{Rice}}\left(P_R;k_2\right)I_{0}\left(k_{2}\Delta\right)=\tilde{\epsilon} %\\
%P_R=F^{-1}_{\mathrm{TWDP}}(\epsilon) \approx F^{-1}_{\mathrm{Rice}}\left(\frac{\epsilon}{I_{0}\left(k_{2}\Delta\right) };k_2\right)  \label{eq:PR_twdp} 
\end{eqnarray}
and the analysis from the Rician case can be directly applied, scaled by $I_{0}\left(k_{2}\Delta\right)$.
From Fig.~\ref{fig:Fig_TWDP} it can be seen that TWDP\footnote{No tractable closed form of PDF or CDF exists. In~\cite{Durgin2002} the PDF is approximated, while we use a complete  expansion as in \cite{saberali2013}. However, due to the numerical sensitivity at URLLC levels, it requires the use of high-precision numerical tools.} starts to differ from a Rician model (with $k_1=k_2$) when $\Delta$ is sufficiently high, such that $\rho_2$ can be distinguished from $V_{\DIF}$. The second specular component  $\rho_2$ lifts-off the lower tail as $\Delta\rightarrow0$ dB, while preserving the Rayleigh tail slope. Note that, in order to reach the extreme slope of the singular TW model at the URLLC levels, one needs $\Delta=0$dB and $k_2$ in range 50 to 60dB, which is very unlikely to happen in practice.

\begin{figure}
\centering
\includegraphics[scale=0.6]{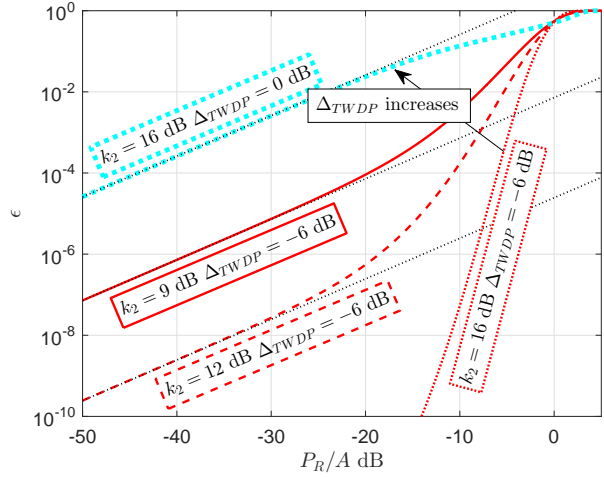} 
\caption{Two-Wave Diffuse Power CDF and its tail approximation (black dotted line): 
$\rho=1$, while $\rho_2$ is set to correspond to the $\Delta$ \eqref{DeltaDefinition} given at each curve. The corresponding $k_2$ \eqref{eq:k-factor} is also given.
}
\label{fig:Fig_TWDP}
\end{figure}

\subsubsection{Suzuki Channel (Suz)}
This is a compound channel consisting of a diffuse component only, which is a mixture between a Rayleigh envelope and a log-normal varying mean
\cite{RVJBA}. The compound envelope is $r=|X_R+jX_I|$, where $X_R$ and $X_I$ are zero-mean Gaussian variables with variance $\sigma_{\mathrm{LN}}=e^{\mathcal N}$ that has a log-normal distribution. 

The PDF and CDF of the Suzuki channel can be found as follows. Let us denote by $A$ the average power used to generate Rayleigh-faded power level $A$. The power $A$ is log-normal distributed, such that we can obtain its PDF from the PDF of the log-normal envelope (\ref{eq:p_LN}) by substituting $A=r^2$. This leads to the following joint distribution of $P$ and $A$:
% * <pe@es.aau.dk> 2017-12-01T18:10:50.765Z:
%
% ^.
\begin{equation}\label{eq:jointPDFSuzuki}
f_{\mathrm{Suz}}(P,A)=\frac{1}{A}e^{-\frac{P}{A}} \cdot \frac{1}{2A\sigma_l\sqrt{2\pi}}e^{-\frac{\left(\frac{1}{2} \ln A -\mu_l \right)^2}{2 \sigma_l^2}}
\end{equation}
For given $P_R$, the outage probability can be calculated as follows:
\begin{equation} \label{eq:IntegralSuzuki}
\epsilon=\int_{0}^{P_R} \mathrm{d}P \int_{0}^{\infty} f_{\mathrm{Suz}}(P,A) \mathrm{d}A
\end{equation}
%The distribution $f_{\mathrm{Suz}}(P,A)$ can be rewritten as follows:
%\begin{equation}\label{eq:PDFSuzuki}
%f_{\mathrm{Suz}}(P,A)= \frac{1}{2A^2\sigma_l\sqrt{2\pi}}
%e^{-\frac{\left(\frac{1}{2} \ln A -\mu_l \right)^2}{2 \sigma_l^2}-\frac{P}{A}}
%\end{equation}

The upper bound for (\ref{eq:IntegralSuzuki}) is obtained by noting that $\frac{P}{A} \geq 0$ and it can be removed from (\ref{eq:jointPDFSuzuki}), after which we get:
\begin{eqnarray}\label{eq:epsilon_Suz_upperbound}
\epsilon &\leq& \int_{0}^{P_R} \mathrm{d}P \int_{0}^{\infty} \frac{1}{2A^2\sigma_l\sqrt{2\pi}}e^{-\frac{\left(\frac{1}{2} \ln A -\mu_l \right)^2}{2 \sigma_l^2}} \mathrm{d}A \nonumber \\
&=& e^{2\sigma_l^2-2\mu_l} P_R = \frac{P_R}{A_{\mathrm{Suz}}} \cdot e^{4\sigma_l^2}
\end{eqnarray}
where $A_{\mathrm{Suz}}=e^{2\sigma_l^2+2\mu_l}=A_{\mathrm{LN}}$ \cite{withers2012}\footnote{Note that $A_{\mathrm{Suz}}=E[r^2]e^{2\sigma_l^2+2\mu_l}$, where $E[r^2]=2\cdot\sigma_{Rayl}^2$ is the mean power of the Rayleigh part. \cite{withers2012} has used unit variance Gaussians $\sigma^2_{Rayl}=1$. Without loss of generality we assume $\sigma_{Rayl}=1/\sqrt{2}$, such that all mean level shifts are attributed solely to $\mu_l$.}.

The lower bound can be found by using the inequality $e^{-x} \geq 1-x$ which leads to:
\begin{equation}
f_{\mathrm{Suz}}(P,A) \geq \frac{1}{2A^2\sigma_l\sqrt{2\pi}}
e^{-\frac{\left(\frac{1}{2} \ln A -\mu_l \right)^2}{2 \sigma_l^2}}(1-\frac{P}{A})
\end{equation}
and results in 
\begin{equation}\label{eq:epsiolon_Suz_lowerbound}
\epsilon \geq \frac{P_R}{A_{\mathrm{Suz}}} e^{4\sigma_l^2}-\left(\frac{P_R}{A_{\mathrm{Suz}}}\right)^2 \cdot e^{12\sigma_l^2} \approx \frac{P_R}{A_{\mathrm{Suz}}} e^{4\sigma_{\mathrm{dB}}^2 \tfrac{\ln(10)^2}{20^2}} = \tilde{\epsilon}
\end{equation}
For URLLC-relevant levels it is $P_R \ll A_{\mathrm{Suz}}$, such that the upper bound can be treated as tight. The tail has a Rayleigh-like slope of $\beta=1$, but pushed to lower levels as seen in Fig. \ref{fig:Fig_shadowed}.

\begin{figure}
\centering
\includegraphics[scale=0.6]{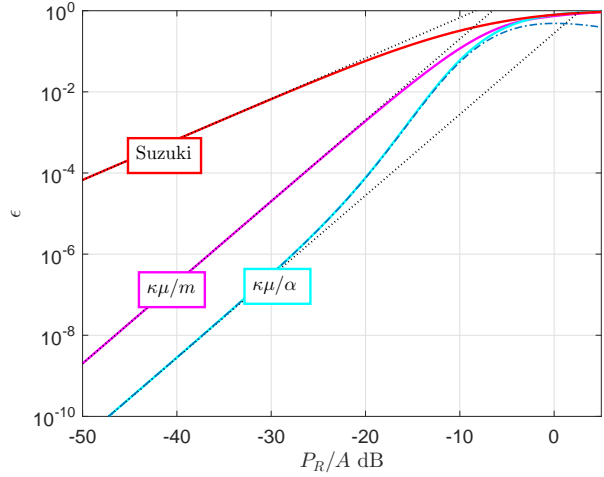} 
\caption{Shadowed CDFs from \eqref{eq:e_kappamu_nak} and numerical integration of \eqref{eq:IntegralSuzuki},\eqref{eq:f_kappamu_IG}  and their tail approximations \eqref{eq:epsiolon_Suz_lowerbound},\eqref{eq:e_kappamu_nak_tail},\eqref{eq:e_kappamu_IG2} (black dotted lines) and heuristic CDF expansion \eqref{eq:e_kappamu_IG_heuristic} (dash-dot curve): Suzuki ($\sigma_{dB}=6$dB), $\kappa-\mu/m$ and $\kappa-\mu/\alpha$  ($\kappa=3.9, \mu=2, m=0.25, \alpha=1.5$). %\MA{@PATRICK all necessary descriptive params here.}
}
\label{fig:Fig_shadowed}
\end{figure}

\subsubsection{Nakagami-m shadowed $\kappa-\mu$ Channel ($\kappa\mu/m$)}

Shadowing the total signal has been investigated in \cite{Yoo2016},\cite{Yoo2015}, but provides complicated PDF and no known closed-form solution for the CDF.
A model that considers shadowing of only the dominant signal parts has been developed by \cite{Paris2014}.
The instant power is $p={\sum_{i=1}^\mu (X_{Ri}+\xi p_i)^2+(X_{Ii}+\xi q_i)^2}$, where %\MA{$\xi$ (I WOULD SUGGEST TO AVOID USING THE SAME SYMBOL FOR DIFFERENT QUANTITIES. ANY SUGGESTION PATRICK?)} \textbf{[well $\xi$ was used in original paper..But we have other clashes : $\Delta$ in TW/TWDP vs curve difference..$\beta$ as slope vs Wei and inv Gamma param..$\alpha$ power law scale vs invGamma param..$\Gamma$ function or correlation coefficient or reflection coef...etc . so either allow 'double symbolism' or big cleaning job all over the paper]} 
$\xi$ is a power normalized Nakagami-m distributed shadowing amplitude acting on specular components $p_i+j q_i = \rho_ie^{j\phi_i}$.
The closed form PDF of $P_R$ (with $p={P_R}/{A_{\mathrm{\kappa\mu/m}}}$) is given by \cite{Paris2014}: 
\begin{align}\label{eq:f_mukappa_nak}
f _{\mathrm{\kappa\mu/m}}(P_R)=\frac{\mu^{\mu}}{\Gamma(\mu)A_{\mathrm{\kappa\mu/m}}} p^{\mu-1} e^{-(1+\kappa)\mu \cdot p}\left(\frac{m}{\kappa\mu+m}\right)^m {}_1F_1\left(m;\mu;\frac{\kappa(1+\kappa)\mu^2}{\kappa\mu+m}p \right)
\end{align}
with $_1F_1(\cdot)$ Kummer's function of the first kind \cite{gradshteyn_ryzhik}. Essentially, the first part is a Nakagami-m PDF of order $\mu$, while the latter part holds a $\mu-$order modified 
Rician impact. }
The tail is:
%given by:
\begin{align}\label{eq:e_kappamu_nak}
\epsilon=F_{\mathrm{\kappa\mu/m}}(P_R) & = \underbrace{ \frac{\mu^{\mu}}{\Gamma(\mu+1)} \left(\frac{P_R}{A_{\mathrm{\kappa\mu/m}}}\right)^{\mu}}_{\approx F_{Nak}(p;\mu)}  (\kappa+1)^\mu \underbrace{\left(\frac{m}{\kappa\mu+m}\right)^m}_{\rightarrow e^{-\kappa\mu}\vert m \rightarrow \infty} \Phi_2(b_1,b_2;c,x,y)\\\label{eq:e_kappamu_nak_tail} 
& \approx  \frac{\mu^{\mu}(\kappa+1)^\mu}{\Gamma(\mu+1)} \left(\frac{P_R}{A_{\mathrm{\kappa\mu/m}}}\right)^{\mu}  \left(\frac{m}{\kappa\mu+m}\right)^m=\tilde{\epsilon}
\end{align}
where $\Phi_2(\cdot)$ is Humberts function \cite{Choi2015} with arguments $(b_1=\mu-m,b_2=m,c=\mu+1,x=-\mu(1+\kappa)\frac{P_R}{A_{\mathrm{\kappa\mu/m}}},y=x\frac{m}{\kappa\mu+m})$.
The power law tail approximation follows from \cite[eq. (13)]{Paris2014}.
%Some bounds on $\Phi_2(\cdot)$ appear in \cite{Joshi1982}, while a numerical calculation procedure is given in \cite{Martos}.
Since $\xi\rightarrow 1$ when $m \rightarrow \infty$, expression \eqref{eq:e_kappamu_nak} should reduce to the regular $\kappa-\mu$ in \eqref{eq:e_kappamu}; this is indeed so, as $\left(\frac{x+m}{m}\right)^m \rightarrow e^x$  and \eqref{eq:e_kappamu_nak} is evidently in accordance with \eqref{eq:e_kappamu}. %  but this is not immediately visible as \eqref{eq:e_kappamu} holds an exponential term not explicit in \eqref{eq:e_kappamu_nak}.
%However, recognizing that $\left(\frac{x+m}{m}\right)^m \rightarrow e^x$ for $m \rightarrow \infty$, accordance with the more \textbf{XXXinformative?XXX} form in \eqref{eq:e_kappamu} appears. 
Fig. \ref{fig:Fig_shadowed} shows a strongly shadowed example ($m=0.25$).
It can be noticed that the shoulder is significantly broadened to a degree that the elevation of the shoulder %'hump' 
visible in the regular $\kappa-\mu$ case, has vanished. This is expected as only the LOS part has been shadowed, thereby effectively averaging %'modulating'
the $\kappa-\mu$ distributions shape 
($\kappa$-factor). Consequently, the tail is being pushed to significantly lower outages.%\PP{The use of 'shoulders', 'humps' and similar is not very precise. Coudl you please try to rephrase your explanation of the figure?}\PE{'shoulders' should be established term..'hump' probably not.. so something like:'elevated shoulder compared to the tail'}

%tohere
\subsubsection{Inverse $\Gamma-$shadowed $\kappa-\mu$ Channel ($\kappa\mu/\alpha$)}
%\MA{@Patrick: I'm missing a little bit of story here...}\textbf{[The idea was Petars SUzuki discussion wrt shadowing vs block fading etc .. should become a general intro/argumentation up front, for all three shadowed distr. = so not have to discuss to much individually]}
Shadowing the total $\kappa-\mu$ (\ref{eq:f_mukappa}) fading envelope $r_{\kappa\mu}=\vert V_{\kappa\mu}\vert$  by an inverse Gamma ($\Gamma^{-1}$) distributed varying mean power $\omega=A_{\kappa\mu}$, leads to a closed form PDF but no tractable closed-form CDF solution \cite{Yoo2015}\footnote{Very recently \cite[(3.14)]{Yoo2017thesis} has provided a closed-form CDF. However, this is provided through a complex Kamp\'{e} de F\'{e}riet function, for which no readily available numerical evaluation exists in tools such as Matlab$^{\mathrm{TM}}$. Furthermore, no simple analytical approximation seems available to be used in an URLLC setting.  The underlying inverse Gamma PDF in \cite[(3.14)]{Yoo2017thesis} seems normalized in a skewed manner with ${\overline{\omega}}=\frac{\alpha}{\alpha-1}$. Thus we make our approximation analysis based on the PDF in \cite{Yoo2015}.}. The combined signal ($r=r_{\kappa\mu}\sqrt{\omega}$) PDF is obtained \cite[(6)]{Yoo2015} by averaging
the conditional envelope PDF $f_{\kappa\mu \vert A_{\kappa\mu}}$ (\ref{eq:f_mukappa}) over the mean power statistics $f_{\Gamma^{-1}}(\omega)=\frac{\beta^{\alpha}}{\Gamma(\alpha)}\frac{1}{\omega^{\alpha+1}}\cdot\mathrm{exp}\left(-\frac{\beta}{\omega} \right)$, with shape $\alpha>0$ and scale $\beta>0$ parameters. The combined signal power PDF of \cite[(10)]{Yoo2015} can be written as
\begin{equation}\label{eq:f_kappamu_IG}
f_{\kappa\mu/\alpha\beta}(P_R)= \frac{(e^{-k}/\kappa\mu )^{\mu}}{B(\alpha,\mu)\left(c \cdot p + 1 \right)^{\alpha-1}}\cdot \underbrace{\frac{ \kappa\mu \cdot c/A_{\kappa\mu/\alpha\beta}}{\left(c \cdot p + 1 \right)^{2}}}_{dx/dP_R} x^{\mu-1} {}_1F_1\left(\alpha+\mu;\mu;x\right)
\end{equation}
with $B(\cdot,\cdot)$ the beta function and argument scaling $c=\frac{\mu(1+\kappa)}{\beta}$ in $x=\kappa\mu\frac{c \cdot p}{ c \cdot p+1}$. The relative power is $p=\frac{r^2}{E[r^2]}=\frac{P_R}{A_{\kappa\mu/\alpha\beta}}$ and $A_{\kappa\mu/\alpha\beta}$ is the mean power of the combined signal. For lower tail levels %$\frac{p}{A_{\kappa\mu/\alpha\beta}} 
$p\ll \frac{1}{c}$ or $\alpha \rightarrow 1^+$, $(c \cdot p +1)^{\alpha-1} \rightarrow 1$ 
%the denominator in the middle term and argument of $1F_1$ in (\ref{eq:f_kappamu_IG}) approaches unity
 and constraining this approximation to the leading term only, we essentially have a function of form $f(x)=x^{b-1}{}_1F_1(a,b,x)$. To obtain the CDF we make use of $\int f(x) dx = x^b\frac{\Gamma(b)}{\Gamma(b+1)}{}_1F_1(a,b+1,x)$ \cite{Wolfram_1F1}.
Thus, via variable transform and reordering  of terms, we arrive at the following upper bound for the CDF:
\begin{equation}\label{eq:e_kappamu_IG}
\epsilon=F_{\kappa\mu/\alpha\beta}(P_R) \leq  \frac{e^{-\kappa\mu}}{\mu B(\alpha,\mu)}  \left(\frac{c \cdot p}{c \cdot p +1} \right)^{\mu} \cdot {}_1F_1\left(\alpha+\mu;\mu+1;\kappa\mu \frac{c \cdot p}{c \cdot p +1}\right)
\end{equation}
which is exact in the limit $\alpha \rightarrow 1^+$. Furthermore, we can simplify (\ref{eq:e_kappamu_IG}) as $_1F_1(a,b, x)\rightarrow 1$ for $x\rightarrow 0$ and realizing that the scale $\beta$ in \cite{Yoo2015} is set arbitrarily, such that $f_{\Gamma^{-1}}$ is not normalized. As $\overline{\omega}=E[\omega]=\frac{\beta}{\alpha-1}$ valid for $\alpha>1$ \cite{Yoo2015}, normalizing shadowing by setting $\overline{\omega}=1$,  we get $\beta=\alpha-1$. Thus, we can represent the impact of the shadowing through a single parameter:
\begin{equation}\label{eq:e_kappamu_IG2}
F_{\kappa\mu/\alpha}(P_R) \gtrsim \underbrace{\frac{(\mu(1+\kappa)e^{-k})^{\mu}}{\Gamma(1+\mu)}  p^{\mu} }_{\approx F_{\kappa\mu}(p;\kappa;\mu)} \frac{\Gamma(\alpha+\mu)}{(\alpha-1)^{\mu}\Gamma(\alpha)} = \tilde{\epsilon},
\end{equation}
i.e. in form of a scaled $\kappa-\mu$ tail, representing a lower bound.
%\textbf{[well at least for kappa  $> 1$ or so]}. 
However, for $\alpha \lesssim 1$ other normalization methods must be used.

We can heuristically reintroduce the denominator term $(c \cdot p +1)^{\alpha-1}$ into the leading term of (\ref{eq:e_kappamu_IG}) for larger arguments: 
\begin{equation}\label{eq:e_kappamu_IG_heuristic}
F_{\kappa\mu/\alpha}(P_R) \gtrsim   \frac{e^{-\kappa\mu}}{\mu B(\alpha,\mu)(c \cdot p +1)^{\alpha-1}}  \left(\frac{c \cdot p}{c \cdot p +1} \right)^{\mu} \cdot {}_1F_1\left(\alpha+\mu;\mu+1;\kappa\mu \frac{c \cdot p}{c \cdot p +1} \right) =\tilde{\epsilon}
\end{equation}where we can redefine $c=\frac{\mu(1+\kappa)}{\alpha-1}$ via the above normalization. This result provides a significantly better fit
%\footnote{tuning the $-1$ in the $(\alpha-1)$ exponent, can improve fit further when $\alpha \gtrsim 2$. A value around $-1.25$ seem to work well.}
than (\ref{eq:e_kappamu_IG}) or (\ref{eq:e_kappamu_IG2}), especially in a strongly shadowed Rician regime ($\kappa/\alpha \gtrsim 1$), as it can be seen in Fig. \ref{fig:Fig_shadowed}. Furthermore, this expanded expression seems to act as a lower bound. It is also seen that the elevated shoulder %'hump'
from the underlying $\kappa-\mu$ signal is better preserved than in the case of the previous $\kappa\mu/m$ model, while also having strong shadowing ($\alpha=1.5$) that indicates that the  complete signal has been shadowed.
%Furthermore, we get
%\begin{equation} \label{eq:log_e_kappamu_IG}
%\log \epsilon \approx \mu \left( \log(p) + \log %\frac{\mu(1+\kappa)e^{-k}}{(\alpha-1)}   \right) - %\log (\mu B(\alpha,\mu)) 
%\end{equation}
%and invert (\ref{eq:e_kappamu_IG2}) to
%\begin{equation}\label{eq:PR_kappamu_IG}
%P_R = F^{-1}_{\mathrm{\kappa\mu/\alpha}}(\epsilon)\approx  \frac{A_{\kappa\mu/\alpha\beta}}{\mu(1+\kappa)e^{-k}} \sqrt[\mu]{\epsilon B(\alpha,\mu) \mu}
%\end{equation}
%\textbf{INCL FIGURE KAPPAMU, NAKAGAMI, WEIBULL and discuss}

%\begin{figure}
%\includegraphics[width=8.3cm]{Fig_kappa_mu_nak_wei.eps} 
%\caption{Weibull, Nakagami and $\kappa-\mu$ distributions following (\ref{eq:e_wei}),(\ref{eq:e_approx_nak}),(\ref{eq:cdf_kappamu}),(\ref{eq:e_kappamu_nak}) and num. integration of (\ref{eq:f_kappamu_IG}). Dotted lines are power law tails (\ref{eq:tail_app_wei}),(\ref{eq:e_nak}),(\ref{eq:e_kappamu}),(\ref{eq:e_kappamu_nak}),(\ref{eq:e_kappamu_IG2}). Dotted curve: heuristic tail (\ref{eq:e_kappamu_IG_heuristic}). Solid dots: 1dB deviation tail approx. vs. distributions}
%\label{fig:kappamu_nak_wei}
%\end{figure}

\section{Other Channels}
\label{sec:other_channels}
In this section we analyze two special models that do not exhibit power-law tail behavior and derive their corresponding tail approximations. 
First, we consider the log-normal distribution \cite{RVJBA}, as a classical reference distribution for shadowing. Next, we treat cascaded channel type models that arise in NLOS propagation, backscatter communication and in 'pin hole' channels \cite{RVJBA}. The two models also represent two extremes, the macro scale (log-normal shadowing) and short range (e.g. device-to-device). 
Furthermore, %the inclusion of 
these models %contribute to the 
can be used as instances to illustrate cases that do not follow the power law in the diversity analysis presented in the next section.

\subsection{Log-Normal Channel (LN)}
\label{sec:Lognormal}

In this model there is a single specular component $N=1$ and no
diffuse component. The specular component is not constant, but subject to a log-normal shadowing, such that log-envelope $\ln(r)$ is modeled as Gaussian variable 
%\cite{Jakes}
\cite{RVJBA}
with PDF:
\begin{equation}\label{eq:p_LN}
f_{\mathrm{LN}}(r)=\frac{1}{r}\mathcal{N}_{\ln(r)} \left(\mu_{l},\sigma_{l}\right)=\frac{1}{r\sigma_l\sqrt{2\pi}}e^{-\frac{\left(\ln(r)-\mu_{l}\right)^{2}}{2\sigma_{l}^{2}}}
\end{equation}
with logarithmic mean and standard deviation \(\mu_l=E[\ln(r)]=\mu_{\mathrm{dB}} \frac{\ln \left(\mathrm{10}\right)}{20}\) and \(\sigma_l=\sqrt{E[\ln(r)^2]-\mu_l^2}=\sigma_{\mathrm{dB}} \frac{\ln \left(\mathrm{10}\right)}{20}\). The average power is $A_{\mathrm{LN}}=e^{2\sigma_l^2+2\mu_l}$ \cite{RVJBA} and the CDF
\begin{equation}\label{eq:e_LN}
\epsilon=F_{\mathrm{LN}}(r)=\tfrac{1}{2}+\tfrac{1}{2}\text{erf}(x(r))
\end{equation}
with $x=(\ln(r)-\mu_l)/(\sigma_l\surd2)$  and erf  being the error function.
Using  B\"urmann-type asymptotic approximation \cite{schoepf2014} leads to 
$F_{\mathrm{LN}}(x) \approx  \tfrac{1}{2}\left(1+\mathrm{sgn}(x) \sqrt{1-e^{-x^2}}\right) \approx \frac{1}{4}e^{-x^{2}}$, when omitting higher order terms and approximating the square root for $|x|\gg 0$.  A tighter approximation can be obtained if we use $F_{\mathrm{LN}}(x) \approx \frac{1}{4}e^{-f\left(x\right)}$ with a polynomial fitting function $f(x)$ %\cite{hamaker1978}
\cite{winitzki}. Comparing $\frac{1}{4}e^{-x^{2}}$ with (\ref{eq:e_LN}), it appears to be shifted proportionally to $\sigma_l$, such that:  
\begin{equation}\label{eq:e_approx_LN}
\epsilon=F_{\mathrm{LN}}\left(P_R\right)\approx\tfrac{1}{4}e^{-\frac{\left(\frac{1}{2}\ln (P_R)-a\sigma_{l}-\mu_{l}\right)^{2}}{2\sigma_{l}^{2}}}=\tilde{\epsilon}
\end{equation}
\begin{equation}\label{eq:log_e_approx_LN}
\log \epsilon \approx \log \tfrac{1}{4}-\frac{1}{\ln(10)}{\frac{\left(P_{R,\mathrm{dB}}-a\sigma_{\mathrm{dB}} -\mu_{\mathrm{dB}} \right)^{2}}{2\sigma_{\mathrm{dB}} ^{2}}}
\end{equation}

With $a=0.223$, the relative error is 
$\eta\lesssim 10^{-1}$
for $10^{-12} \leq\epsilon\leq 10^{-2}$  and $3 \leq\sigma_{\mathrm{dB}}\leq 24$dB. The deviation on the margin matters most for outage analysis and is here below $\tfrac{1}{3}$dB. This accuracy is still very useful, considering the simplicity of the expression for analytical studies.

Finding a root of the second order equation in $\ln(P_R)$ \eqref{eq:e_approx_LN} we get
\begin{equation}
P_R=F^{-1}_{\mathrm{LN}}(\epsilon) \approx e^{2[(a\sigma_l+\mu_l)+\sqrt{2} \sigma_l \sqrt{-\ln(\epsilon)+\ln(1/4)}]}
\end{equation}

For a given $P_R$, we can find the log-log slope as $\beta=\frac{d\log\left(F_{LN}\right)}{d\log\left(P_R\right)}\approx \frac{10}{\ln10} \left[\frac{a}{\sigma_{\mathrm{dB}}}-2\frac{P_{R,\mathrm{dB}}-\mu_{\mathrm{dB}}}{2\sigma_{\mathrm{dB}}^{2}} \right]$. %Clearly, the log-log relationship is not linear as we have seen in the other models. 
From Fig.~\ref{fig:cascade_ral_lognorm} it is observed that, for large $\sigma_{\mathrm{dB}}$, a log-normal channel can exhibit extreme slopes when the level $\frac{P_R}{A}$ is in the region $-10$ to $-30$ dB, which makes it hard to distinguish from a TW or TWDP channel. However, when going towards URLLC-relevant levels, the deviation from a linear slope is noticeable. 
%\PP{The log-normal model is distinctive in sense, the probability of outage is calculated based on a block fading model as discussed
%in subsection \ref{sec:generalizations}. - I DID NOT GET THIS PART, ARE NOT ALL CALCULATED BASED ON A BLOCK FADING?}\PE{well this is originally your distinction between the short term models and the shadow model (stuff w assume sporadic transmission etc).. in later iteration that discussion text was just moved up in front (to also include the shadow mix distributions)}

\subsection{Cascaded Rayleigh Channel (Cas)}
This model also contains only a diffuse component, which is a product of the envelopes of two Rayleigh links $r_1$ and $r_2$. The compound received envelope is $r=r_1r_2=|X_{R_1}+jX_{I_1}|\cdot|X_{R_2}+jX_{I_2}|$ with PDF equal to \cite{simon2005digital,chau2012second}\footnote{Ref. \cite{chau2012second} uses the convention $\Gamma=\Gamma_{V_1,V_2}$, whereas we use $\Gamma=\Gamma_{P_1,P_2}$ as in \cite{simon2005digital}. Note that for Rayleigh fading $|\Gamma_{V_1,V_2}|^2=\Gamma_{X_{R_1},X_{R_2}}^2+\Gamma_{X_{R_1},X_{I_2}}^2=\Gamma_{P_1,P_2}\approx \Gamma_{r_1,r_2}$ and $\Gamma_{X_{R_1},X_{I_2}}^2 \approx 0$ for random links %\cite{kerr1951propagation,
\cite{pierce1960multiple}.}: 
 \begin{equation} \label{eq:p_cascade}
f_{\mathrm{Cas}}(r)=\frac{r_{\Gamma}}{\sigma_{1}\sigma_{2}}I_{0}\left(r_{\Gamma}\sqrt\Gamma\right)K_{0}\left(r_{\Gamma}\right)
\end{equation} 
where $r_{\Gamma}=\frac{r}{\sigma_{1}\sigma_{2}\left(1-\Gamma\right)}$. Using \eqref{eq:p_cascade} we get  \(A_{\mathrm{Cas}}=E[r^2]=4\sigma_1^{2}\sigma_2^{2}(1+\Gamma)=\bar{P_1}\bar{P_2}(1+\Gamma)\) with correlation coefficient $\Gamma$ between powers \(P_1=r_1^2\) and \(P_2=r_2^2\). $I_n$  and $K_n$ are the Modified Bessel functions of 1$^\text{st}$ and 2$^\text{nd}$ kind, of order $n$. The CDF follows as:

\begin{equation}\label{eq:e_cascade}
\epsilon=F_{\mathrm{Cas}}(P_R)=1-r_{\Gamma}[\sqrt\Gamma\cdotp I_{1}\left(r_{\Gamma}\sqrt\Gamma\right)K_{0}\left(r_{\Gamma}\right) +I_{0}\left(r_{\Gamma}\sqrt\Gamma\right)K_{1}\left(r_{\Gamma}\right)]
\end{equation}

Approximating the Bessel functions for $r_{\Gamma}\ll1$, the general case ($\Gamma<1$)  simplifies as
\begin{multline} \label{eq:e_approx_cascade}
 \epsilon =F_{\mathrm{Cas}}(P_R) \approx-\frac{r_{\Gamma}^{2}}{4}\left(1-\Gamma\right)\left[2\ln\left(\frac{r_{\Gamma}}{2}\right)+\left(2\gamma-1\right)\right]
 \\ \approx- \frac{P_R}{A_{\mathrm{Cas}}} \frac{1+\Gamma}{1-\Gamma}{\ln\left(\frac{P_R}{A_{\mathrm{Cas}}} \frac{1+\Gamma}{(1-\Gamma)^2}\right)}=\tilde{\epsilon}
\end{multline}
where $\gamma=0.5772..$ is Euler's constant. The log probability is:
\begin{equation} \label{eq:log_e_approx_cascade}
\log(\tilde{\epsilon}) = \log \frac{P_R}{A_{\mathrm{Cas}}}+\log\frac{1+\Gamma}{1-\Gamma}+\log \left( -\ln \left( \frac{P_R}{A_{\mathrm{Cas}}} \frac{1+\Gamma}{(1-\Gamma)^2}\right)\right)
\end{equation}
and is valid below the knee point
$\frac{P_R}{A_{\mathrm{Cas}}} \ll\frac{1}{4}\frac{\left(1-\Gamma\right)^2}{1+\Gamma}$.  
The slope is found as $\beta \approx \frac{d \log(\tilde{\epsilon})}{d \log \frac{P_R}{A_{\mathrm{Cas}}}}$,
%the first derivative of $\log(\tilde{\epsilon})$ w.r.t. $\log \frac{P_R} {A_{\mathrm{Cas}}}$,
leading to $\beta \approx 1+\frac{1}{\ln \left( \frac{P_R}{A_{\mathrm{Cas}}} \right)+\ln\frac{1+\Gamma}{(1-\Gamma)^2}}$ and it gradually approaches a Rayleigh slope for $\frac{P_R}{A_{\mathrm{Cas}}}\rightarrow 0$.
For $\Gamma=0$ the model collapses to the well known double-Rayleigh model~\cite{JBAIZK}.

When $\tilde{\epsilon} <1/e$, the function in 
(\ref{eq:e_approx_cascade}) can be inverted \cite{Maple}
via a single Lambert W branch \cite{corless1996}:
\begin{equation}
P_R = F^{-1}_{\mathrm{Cas}}(\epsilon) \approx -A_{\mathrm{Cas}} \tilde{\epsilon} \frac{1-\Gamma}{1+\Gamma}\frac{1}{W_{-1}\left(-\frac{\tilde{\epsilon}}{1-\Gamma}\right)}
\end{equation}

For the singular case of $\Gamma=1 (r_1=r_2)$, simple deduction yields $r_{\mathrm{Cas}}=r_1r_2=F_{\mathrm{Cas}}^{-1}(\epsilon) = F_{\mathrm{Rayl}}^{-1}(\epsilon)^2=r_{\mathrm{Rayl}}^2$. Thus, $F_{\mathrm{Cas}}(P_R)=F_{\mathrm{Rayl}}(\sqrt{P_R}) \sim \sqrt{\frac{P_R}{A_{\mathrm{Rayl}}}}$ and the slope $\beta \approx \tfrac{1}{2}$ is identical to the singular case of a TW model. It can be concluded from  Fig.~\ref{fig:cascade_ral_lognorm} that a the log-log behavior of cascaded Rayleigh fading can be represented by two different slopes with a breakpoint.
\begin{figure}
\centering
\includegraphics[width=8.3cm]{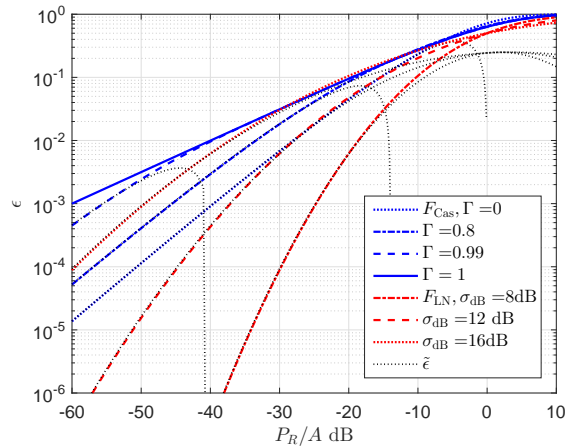} 
\caption{Cascaded Rayleigh (\ref{eq:e_cascade}) with equal power links ($\sigma_1=\sigma_2=\sigma$) - and Log-Normal (\ref{eq:e_LN}) distributions, for different values of the parameters. Dotted curves: tail approx. (\ref{eq:e_approx_cascade}), (\ref{eq:e_approx_LN}).% Solid dots: 1dB deviation tail approx. vs. distributions.
}
\label{fig:cascade_ral_lognorm}
\end{figure}

\section{Simplified Analysis of Diversity Schemes}
%So far we have determined the URC-relevant statistics of various channel models by assuming a single antenna at the receiver. 
In practice, attaining very high reliability levels with reasonable power can only happen by having high levels of diversity at the receiver. Our analysis has shown that the tail approximation at the URLLC levels mostly has the form given in (\ref{eq:MainResult}), which 
can be used for simplified diversity analysis, in particular in cases in which the full PDF/CDF are not tractable. 
%In the previous sections our analysis has focused chiefly on $\beta$.
%, while the scaling $\alpha$ need to be extracted from the tail expressions explicitly. 
%The factor $\alpha$ can be composed of two factors: (1) the tail offset of the distribution $\alpha_\mathrm{CDF}$ determined by the distribution of the channel model; (2) $\alpha_\mathrm{ant}$ due to branch power ratio (BPR), defined for a pair of the $m-$th and $n-$th antenna as \PE{$\mathrm{BPR_{mn}}=\tfrac{A_m}{A_n}$}. Hence, $\alpha=\alpha_\mathrm{CDF}\alpha_\mathrm{ant}$. 

For small terminals, the main impairment towards exploiting multi-antenna (multi-branch) diversity is the branch power ratio (BPR), defined for a pair of the $m-$th and $n-$th antenna as $\mathrm{BPR_{mn}}=\tfrac{A_m}{A_n}$ \cite{plicanic2009,yanakiev2012}.
%,nielsen2014}
On the other hand, when operating in higher frequency bands, the terminal dimensions become comparable or larger than the wavelength, which leads to low branch correlations. Therefore, in the following we assume that the receiver has $M$ antennas that are not correlated, i.e., the received signals across antennas are independent non-identically distributed (i.n.i.d.) random variables (RV).

In Selection Combining (SC), only the strongest signal among the $M$ antennas is selected:
\begin{equation}\label{eq:SC_r}
P_{R,\mathrm{SC}}=\max(P_1,..,P_M)
\end{equation}
For independent branches, the CDF can be expressed as a simple product of the individual CDFs across branches:
\begin{equation}\label{eq:F_SC}
\epsilon=F_\mathrm{SC}\left(P_R\right)=\prod_{m=1}^{M}F_{m}\left(P_R\right)
\end{equation}
When Maximum Ratio Combining (MRC) is used, the received power is:
\begin{equation}\label{eq:MRC_PR}
P_{R,\mathrm{MRC}}=\sum _{m=1}^{M}{P_{m}}
\end{equation}
%When considering the 
The simplest distribution we consider, is the Rayleigh case.
%, a general solution for the CDF with MRC is involved. 
For uncorrelated Rayleigh branch signals, tractable expressions for the tail exist when all BPR$=$1 or all BPR$\neq$1
%,Jakes
\cite{RVJBA,schwartz1996communication}. \cite{schwartz1996communication} suggests a simple approximation for the MRC PDF for low powers, valid for all BPRs; using the approximation in \cite{schwartz1996communication} one can readily obtain:
\begin{equation}\label{eq:F_mrc_rayleigh}
\epsilon=F_{\mathrm{MRC}}\left(P_R\right)
\sim  \underbrace{\frac{1}{M!}}_{\alpha_\mathrm{MRC}}
\underbrace{ \prod_{m=1}^{M}\alpha_{m} \frac{P_R}{A_m}}_{\sim F_{\mathrm{SC}}}=  \underbrace{\frac{1}{M!}}_{\alpha_\mathrm{MRC}} \underbrace{\left(\frac{P_R}{A_1} \right)^M \prod_{m=1}^{M}\frac{\alpha_m}{\mathrm{BPR}_{m1}}}_{\sim F_{\mathrm{SC}}}=\tilde{\epsilon}
\end{equation} 
which has  simple product form as in SC, shifted by $\alpha_\mathrm{MRC}=\tfrac{1}{M!}$. 

Next, we derive an approximation of the CDF for general M-branch MRC at low powers (see Appendix \ref{app:App Diversity}); the approximation is of the same form as \eqref{eq:F_mrc_rayleigh}:
%, i.e., it can be expressed as a simple product as follows:
\begin{equation}\label{eq:F_analyt_mrc}
\epsilon=F_\mathrm{MRC}(P_R) \approx \underbrace{\frac{\prod_{m=1}^M \Gamma(1+\beta_m)}{ \Gamma(1+\sum_{m=1}^M \beta_m)}}_{\alpha_\mathrm{MRC}}   \underbrace{\prod_{m=1}^M \alpha_m \left(\frac{P_R}{A_m}\right)^{\beta_m}}_{\sim F_{SC}=\prod F_{m}}=\tilde{\epsilon}
\end{equation}
Note that this solution also splits into a MRC weighting term $\alpha_\mathrm{MRC}$, which depends solely on the branch slopes $\beta$ and correctly collapsing to 1 for $M=1$, and a term similar to SC $F_\mathrm{SC}=\prod F_m$, which involves the offsets $\alpha$.
The distribution specific parameters $\alpha, \beta$ for this simple expression, are given in Table \ref{tab:tail_approx}. Inserting $\alpha_{\kappa\mu/m}$ and $\beta_{\kappa\mu/m}$ from  Table \ref{tab:tail_approx}, does indeed produce the distribution specific MRC solution for shadowed $\kappa-\mu$ fading given in \cite[(18)]{Paris2014}.
When all branch slopes are equal $\beta_m=\beta$, we get:
\begin{equation}\label{eq:F_analyt_mrc_beta}
\epsilon=F_\mathrm{MRC}(P_R) \approx \underbrace{\frac{\Gamma(1+\beta)^M}{ \Gamma(1+\beta M)}}_{\alpha_\mathrm{MRC}}
\underbrace{ \left( \frac{P_R}{  A_{1}} \right)^{\beta M}  \prod_{m=1}^M  \frac{\alpha_m}{ \mathrm{BPR}_m^{\beta}}  }_{\sim F_{SC}=\prod F_m}=\tilde{\epsilon}
\end{equation}
with $\mathrm{BPR}_m=A_{m}/A_{1}$. A heuristic simplification is $\alpha_\mathrm{MRC} \sim \tfrac{1}{M!^\beta}$, with 
outage error $\lesssim1$dB for $M=4$ and $\lesssim1.5$dB for $M=8$, both at $10^{-6}$ probability and for $\tfrac{1}{2}\lesssim\beta\lesssim2$.
For the particular Rayleigh case of $\beta=1$, the solution collapses to the known result of  \eqref{eq:F_mrc_rayleigh}.

Using \eqref{eq:F_analyt_mrc}, we can bound the tail approximation error using the approximation error functions derived previously, resulting in:
\begin{equation}
\phi_{\mathrm{MRC}}(P_R) = (1+\phi^{\max}(P_R))^M - 1,
\end{equation}
with $\phi^{\max}(P_R)=\max(\phi_1(P_R),\hdots,\phi_M(P_R))$.
Using Bernoulli approximation, we arrive at the intuitive expression $\phi_{\mathrm{MRC}}(P_R)\approx M\phi^{\max}(P_R)$, which can be used for quick evaluation of the upper bound on the power $P_R$ for given error tolerance $\eta$.%$\varepsilon$. 

Finally, based on \eqref{eq:F_analyt_mrc}
%Furthermore, as this solution also appears in a weighted form of the SC solution,
it is easy to make a heuristic generic expansion by considering local log-log linear approximation of \emph{any} CDF tail. This is e.g. the case for Log-Normal and Cascaded Rayleigh models \footnote{or some of the more elaborate tails, like for $\kappa-\mu$ type models etc.}, where the branch slopes depend on the power levels $\beta(P_R/A)$, such that:
\begin{equation}\label{eq:F_mrc_generic}
\epsilon=F_\mathrm{MRC}(P_R)\approx \alpha_\mathrm{MRC}(\beta_1(P_R)..\beta_M(P_R)) F_\mathrm{SC}\left(P_R\right)=\tilde{\epsilon}
\end{equation}
with $\alpha_\mathrm{MRC}$ given in \eqref{eq:F_analyt_mrc} or simplified in \eqref{eq:F_analyt_mrc_beta} when all branches have the same log-log slope.
With this structure and availability of slopes $\beta(P_R)$, one can use the full CDFs in $F_{SC}$.%, if the corresponding slopes $\beta(P_R)$ are available.

%{\PE{[Marko can you make small section wrt diversity combined approx errors become multiplicative in above SC terms?]
\begin{figure}
\centering
\includegraphics[width=8.3cm]{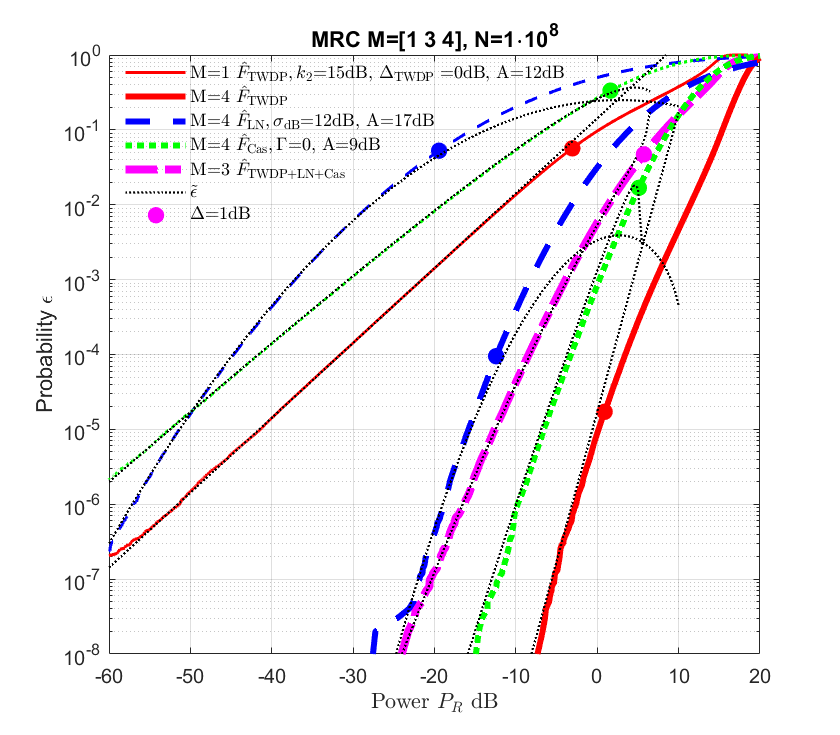} 
\caption{MRC with $M=4$ for i.i.d. TWDP, Log-Normal and Double-Rayleigh (Cascaded Rayleigh with $\Gamma=0$).
%\PE{with uncorrelated signals=WELL i.i.d. MEANS JUST THIS, SO DOUBLE STATEMENT IN A WAY}
Also MRC with $M=3$ for one branch of each of the distributions. Bolder curves are obtained by simulation, thin dotted lines are the tail approximations.   \eqref{eq:e_approx_twdp}, \eqref{eq:e_approx_cascade}, \eqref{eq:e_approx_LN} and MRC \eqref{eq:F_mrc_generic}, \eqref{eq:F_analyt_mrc}. Bold dots represent 1dB deviation of tail vs. simulation.}
\label{fig:MRC}
\end{figure}

Fig. \ref{fig:MRC} shows Monte Carlo simulation with  $10^8$ samples of  TWDP, Log-Normal and Double-Rayleigh (Cascaded Rayleigh with $\Gamma=0$) distributions with different mean powers. Each distribution is further circularly shifted to provide uncorrelated copies for i.i.d. $M=4$. 
It is observed how the single branch tail approximations (thin lines) follow the simulations up to the onset of the shoulders. %For the TWDP approximation, it is  $-k_2$ dB from the mean level.
The $4-$branch MRC tail approximation shows very good fit at URLLC probabilities - bold dots indicate point of 1dB deviation to the simulation.  Furthermore, from the log-normal and 3-branch cross-distribution MRC, it is observed that the heuristic expansion in \eqref{eq:F_mrc_generic} indeed provides very useful results.

\section{Discussion and Conclusions}

We have investigated the properties of wireless channel models in the URLLC regime and developed approximations of the tail distributions. 
Furthermore, our analysis has shown that, for a wide range of practical models, the outage probability at URLLC levels depends on the minimal required decoding power through an exponent $\beta$ which, for the case of Rayleigh fading is $\beta\approx1$. More importantly, it has also been shown  that the outage probability also depends on a tail offset $\alpha$, which is strongly dependent on specific specular and diffuse part combinations.

The previous URLLC studies \cite{Ericsson_URC_2015} have adopted Rayleigh models. Our analysis reveal it is not so much the existence of such a power law outage model with exponent $\beta$, but rather the tail offset $\alpha$, which needs attention in assessing the tail probability.
We have seen that hyper-Rayleigh fading,
can result in an exponent $\beta=\frac{1}{2}$ at URLLC levels, which leads to very conservative, if at all useful, rate selection. Several models
can lead to hyper-Rayleigh fading,
despite the fact that the physical basis of the models are different, such as: cancellation of dominant paths, cascaded links, multiple scattering or shadowing. 
%the first being extremely parameter sensitive  while the latter ones being more robust against parameter variations.
Hence, when conducting empirical studies that work with these models, one should account for the large uncertainty that occurs when assessing the models at the URLLC levels and collect proportionally large number of samples. Finally, we have provided a simplified analysis of MRC diversity for power-law tails, as well as a 
heuristic generic expansion. This paves the way for more elaborate diversity analysis, % of various diversity schemes
which is, on the other hand, vital for achieving URLLC operation with reasonable data rates. 

We have used a comprehensive list of channel models and in this sense the present paper can be considered as a reference work for the channel models in URLLC regime. The necessary next step is to relate the models to the experimental results that are relevant for the URLLC regime. This requires a careful design of the measurement procedures, considering that one needs to capture rare events and thus potentially use a large number of measurement samples. 

\section*{Acknowledgment}
This work was supported by the H2020 European Research Council (ERC Consolidator Grant Nr. 648382 WILLOW).
%\textbf{WRT ref [Jakes]=[7?] you have DC Cox in as editors as well??? I never seen this. Cox is 'only' co-author on one of the chapters}

\begin{appendices}
{\section{Derivation of the Approximation Error Function}
\label{app:App Error functions} 
\textbf{Two-Wave Model:}
We derive the tail for the general case when $\Delta\leq 1$ (which includes $\Delta = 1$ as a special case).
By definition the tail is the solution to the following integral:
\begin{equation}\label{eq:tail_integral_twowave}
\epsilon = \int_{\sqrt{A_{\mathrm{TW}}(1-\Delta)}}^{\sqrt{P_R}}\frac{2r}{\pi A_{\mathrm{TW}}\sqrt{\Delta^2 - \left(1-\frac{r^2}{A_{\mathrm{TW}}}\right)^2}}d r.
\end{equation}
We introduce the following variable $x\equiv x(r)$:
\begin{equation}
x = \sqrt{\frac{1}{\Delta}r^2 - \frac{1-\Delta}{\Delta}A_{\mathrm{TW}}}.
\end{equation}
Using change of variables, the integral \eqref{eq:tail_integral_twowave} can be written as follows:
\begin{equation}
\epsilon = \int_{0}^{\sqrt{P_R^{\star}}}\frac{2x}{\pi A_{\mathrm{TW}}\sqrt{1 - \left(1-\frac{x^2}{A_{\mathrm{TW}}}\right)^2}}d x,
\end{equation}
where $P_R^{\star} = x^2(P_R)$. 
Even though the CDF has a closed form expression, bounding the sum of the higher order terms of its series expansion is difficult.
We use an alternative approach instead.
Specifically, we expand the integrand into Taylor series in the interval $[0,x),x\geq 0,x\rightarrow 0$ using Lagrange form for the remainder (i.e., the sum of the remaining higher order terms):
\begin{equation}\label{eq:Taylor_exact_Lagrange}
f(x) = f(0) + \frac{f'(0)}{1!}x + \frac{f''(\delta)}{2!}x^2,
\end{equation}
for $\delta\in(0,x)$.
Then, we bound the remainder from above, relying on the fact that integrating will not change the inequality; we obtain the following:
\begin{align}
\epsilon & = \int_{0}^{\sqrt{P_R^{\star}}}\left(\frac{1}{\pi}\sqrt{\frac{2}{A_{\mathrm{TW}}}} + \frac{4(A_{\mathrm{TW}} - \delta^2)}{\pi\sqrt{(2A_{\mathrm{TW}} - \delta^2)^3}}x^2\right)dx\\\label{eq:TW_step1}
& \leq \int_{0}^{\sqrt{P_R^{\star}}}\left(\frac{1}{\pi}\sqrt{\frac{2}{A_{\mathrm{TW}}}} + \frac{4(A_{\mathrm{TW}} - P_R^{\star})}{\pi\sqrt{(2A_{\mathsf{TW}} - P_R^{\star})^3}}x^2\right)dx\\
& = \frac{1}{\pi}\sqrt{\frac{2}{A_{\mathrm{TW}}}}\sqrt{P_R^{\star}} + \frac{4(A_{\mathrm{TW}} - P_R^{\star})}{\pi\sqrt{(2A_{\mathrm{TW}} - P_R^{\star})^3}}\frac{\sqrt{(P_R^{\star})^3}}{3}\\\label{eq:TW_step_last}
& = \frac{1}{\pi}\sqrt{\frac{2}{A_{\mathrm{TW}}}}\sqrt{P_R^{\star}}\left(1 + \frac{4}{3}\sqrt{\frac{A_{\mathrm{TW}}}{2}}\frac{(A_{\mathrm{TW}} - P_R^{\star})P_R^{\star}}{\pi\sqrt{(2A_{\mathrm{TW}} - P_R^{\star})^3}}\right),
\end{align}
Recognizing that \eqref{eq:TW_step_last} can be written as $\epsilon\leq\tilde{\epsilon}(1 + \phi(P_R))$, we extract $\phi(P_R)$ as the second term in the brackets in \eqref{eq:TW_step_last}, completing the derivation. 
Note that in \eqref{eq:TW_step1} we used the fact that the multiplicative term in front of $x^2$ increases monotonically with $\delta\in(0,x)$; hence we bound it from above with $x=\sqrt{P_R^{\star}}$.
}

\textbf{Rayleigh Model:}
Deriving the approximation error function follows similar steps as in the TW case, except that we directly bound the Lagrange remainder of the Taylor series expansion of the tail in the interval $[0,P_R),P_R\rightarrow 0$.
Hence, we obtain:
\begin{align}
\epsilon = \frac{P_R}{A_{\mathsf{Rayl}}} - \frac{P_R^2}{A_{\mathsf{Rayl}}^2}e^{-\frac{P_R}{A_{\mathsf{Rayl}}}} \geq \frac{P_R}{A_{\mathsf{Rayl}}} - \frac{P_R^2}{A_{\mathsf{Rayl}}^2} = \frac{P_R}{A_{\mathsf{Rayl}}}\left(1 - \frac{P_R}{2A_{\mathsf{Rayl}}}\right),
\end{align}
which completes the derivation.

\textbf{Weibull Model:}
We use the inversion:
\begin{equation}\label{eq:Wei_step1}
P_R = \frac{A_{\mathrm{Wei}}}{\Gamma(1+1/\beta)}\left(-\ln(1-\epsilon)\right)^{\frac{1}{\beta}},
\end{equation}
and the following bounds:
\begin{equation}\label{eq:Wei_step2}
\epsilon + \frac{\epsilon^2}{(1-\epsilon)}\geq-\ln(1-\epsilon)\geq\epsilon.
\end{equation}
The upper bound in \eqref{eq:Wei_step2} has been derived by bounding from above the remainder of the Taylor expansion of the function $-\ln(1-\epsilon)$ in the interval $[0,\epsilon)$, i.e.:
\begin{equation}
\frac{\epsilon^2}{2(1-\epsilon)^2}\leq \frac{\epsilon^2}{(1-\epsilon)},
\end{equation}
for $\epsilon\geq 0$.
Replacing \eqref{eq:Wei_step2} into \eqref{eq:Wei_step1}, we get:
\begin{equation}
\frac{A_{\mathrm{Wei}}}{\Gamma(1+1/\beta)}\epsilon^{\frac{1}{\beta}}\left(\frac{\epsilon}{1-\epsilon}\right)^{\frac{1}{\beta}}\geq P_R\geq \frac{A_{\mathrm{Wei}}}{\Gamma(1+1/\beta)}\epsilon^{\frac{1}{\beta}},
\end{equation}
which, after inverting for $\epsilon$, can be written as:
\begin{equation}
\tilde{\epsilon}\geq\epsilon\geq\tilde{\epsilon}\frac{1}{1+ \left(\Gamma(1+1/\beta)\frac{P_R}{A_{\mathsf{Wei}}}\right)^{\beta}} = \tilde{\epsilon}\left(1 - \frac{\left(\Gamma(1+1/\beta)\frac{P_R}{A_{\mathsf{Wei}}}\right)^{\beta}}{1+ \left(\Gamma(1+1/\beta)\frac{P_R}{A_{\mathsf{Wei}}}\right)^{\beta}}\right).
\end{equation}

\textbf{Rician, Nakagami-m and $\kappa-\mu$ Model}:
Due to space limitation, we derive the approximation error function only for the general $\kappa-\mu$ model; the corresponding error functions for the Rician and Nakagami-m models can be obtained as special cases.\footnote{Note that they can be derived separately using similar reasoning.}
We use polynomial series expansion for the generalized Marcum Q-function via generalized Laguerre polynomials and write the CDF as follows {\cite{Andras2011}}:
\begin{equation}\label{eq:km_tail_expans}
\epsilon = e^{-\kappa\mu}\sum_{n=0}^{\infty}(-1)^n \frac{L_n^{(\mu-1)}(\kappa\mu)}{\Gamma(\mu+n+1)}\left((\kappa+1)\mu\frac{P_R}{A_{\kappa\mu}}\right)^{n+\mu},
\end{equation}
where $L_n^{(\alpha)}(\cdot)$ is the generalized Laguerre polynomial of degree $n$ and order $\alpha$.
Recognizing that the first term in the above sum gives the power law approximation $\tilde{\epsilon}$, we obtain the following:
\begin{align}
|\epsilon - \tilde{\epsilon}| & = \left|e^{-\kappa\mu}\sum_{n=1}^{\infty}(-1)^n \frac{L_n^{(\mu-1)}(\kappa\mu)}{\Gamma(\mu+n+1)}\left((\kappa+1)\mu\frac{P_R}{A_{\kappa\mu}}\right)^{n+\mu}\right|\\\label{eq:km_step1}
&\leq e^{-\kappa\mu}\sum_{n=1}^{\infty} \frac{|L_n^{(\mu-1)}(\kappa\mu)|}{\Gamma(\mu+n+1)}\left((\kappa+1)\mu\frac{P_R}{A_{\kappa\mu}}\right)^{n+\mu}\\\label{eq:km_step2}
&\leq \frac{e^{-\frac{\kappa\mu}{2}}}{\Gamma(\mu)}\sum_{n=1}^{\infty} \frac{\Gamma(\mu + n)}{n!\Gamma(\mu+n+1)}\left((\kappa+1)\mu\frac{P_R}{A_{\kappa\mu}}\right)^{n+\mu}\\\label{eq:km_step3}
&\leq \frac{e^{-\frac{\kappa\mu}{2}}}{\Gamma(\mu)\mu}\left((\kappa+1)\mu\frac{P_R}{A_{\kappa\mu}}\right)^{\mu}\sum_{n=1}^{\infty} \frac{1}{n!}\left((\kappa+1)\mu\frac{P_R}{A_{\kappa\mu}}\right)^{n}\\
& = \tilde{\epsilon} e^{\frac{\kappa\mu}{2}}\left(e^{(\kappa+1)\mu\frac{P_R}{A_{\kappa\mu}}} - 1\right),
\end{align}
which completes the derivation.
In \eqref{eq:km_step2} we used the following upper bound \cite{Andras2011}:
\begin{equation}
|L_n^{(\alpha)}(x)|\leq\frac{\Gamma(\alpha + n + 1)}{n!\Gamma(\alpha + 1)}e^{\frac{x}{2}},
\end{equation}
and in \eqref{eq:km_step3} we used:
\begin{equation}
\frac{\Gamma(\mu + n)}{\Gamma(\mu + n + 1)} = \frac{(\mu + n - 1)!}{(\mu + n)!}= \frac{1}{\mu + n}\leq \frac{1}{\mu},
\end{equation}
for $n\geq 1$.

%\MA{\textbf{Generic M-branch MRC Combining}:
%Replacing $F_{\mathrm{SC}}(P_R)$ with \eqref{} in \eqref{}, and using \eqref{}, we obtain the following expression:
%\begin{equation}
%\prod_{m=1}^M\tilde{\epsilon}_m(1-\phi_m(P_R))\leq\epsilon\leq\prod_{m=1}^M\tilde{\epsilon}_m(1+\phi_m(P_R))
%\end{equation}
%which can be further rewritten as:
%\begin{equation}
%(1-\phi^{\max}(P_R))^M\prod_{m=1}^M\tilde{\epsilon}_m\leq\epsilon\leq(1+\phi^{\max}(P_R))^M\prod_{m=1}^M\tilde{\epsilon}_m
%\end{equation}
%}

\section{MRC for Random Variables with Power-Law Tails}
\label{app:App Diversity} 

A general M-branch MRC PDF solution for independent RV can be obtained through a convolution of the branch PDFs  $f_1\ast f_2 ..\ast f_M$, e.g. through the multiplication of moment generating functions (MGF) $\prod M_m$ and inverse Laplace transform $\mathcal{L}^{-1}$.  However, approximating the full CDF this way, can result in too complex solutions to readily extract a simple tail approximation. However, it is sufficient to deal with branch tail PDFs only \cite{hitczenko1999}.
The lower tail PDF corresponding to \eqref{eq:MainResult} can be obtained as:
\begin{equation}\label{eq:f_approx}
f(P_R)  \underset{P_R \rightarrow 0}{\approx} \frac{d\alpha \left(\frac{P_R}{A}\right)^{\beta}}{dP_R} =\alpha\left( \frac{1}{A}\right)^{\beta}\beta P_R^{\beta-1} 
\end{equation}
Using Laplace transform relation \cite[ET I 137(1), Table 17.13]{gradshteyn_ryzhik}  $F(s)=\mathcal{L}( f(t) ) =1/s^\nu\leftrightarrow f(t)=t^{\nu-1}/\Gamma(\nu)$, the branch $F(s) \approx \alpha\left( \frac{1}{A}\right)^{\beta}\beta\frac{\Gamma(\beta)}{s^\beta} $. The i.n.i.d M-branch MRC CDF (for any BPR or $\beta$ combination), is established as $F_\mathrm{MRC}(P_R)=\mathcal{L}^{-1} \left(\frac{1}{s} \prod_{m=1}^{M} F_{m}(s)\right)$, where $\frac{1}{s}$ is used to produce the CDF from the inverse transform. 
Using the same Laplace relation as before, we arrive at
\begin{multline}\label{eq:F_analyt_mrc_laplace}
\epsilon=F_\mathrm{MRC}(P_R) \approx \mathcal{L}^{-1} \left(\frac{1}{s \prod_{m=1}^{M} s^{\beta_m} } \right) \cdot \prod_{m=1}^{M} \alpha_m \beta_m \Gamma(\beta_m)  \left( \frac{1}{A_m}\right)^{\beta_m} =\tilde{\epsilon}
\end{multline}
which after reordering of terms appears in the form given in (\ref{eq:F_analyt_mrc}).

\end{appendices}

\bibliographystyle{IEEEtran}

\end{document}